\documentclass[]{scrartcl}
\pdfoutput=1 

\usepackage{amsmath}
\usepackage{amssymb}
\usepackage{algorithm}
\usepackage{algorithmicx}
\usepackage{algpseudocode}
\usepackage{multirow}
\usepackage{booktabs}
\usepackage{colortbl}
\usepackage{url}
\usepackage[switch]{lineno}
\usepackage{epstopdf}
\usepackage{graphicx}

\title{Rank-One Network: An Effective Framework for Image Restoration}
\author{Shangqi~Gao and~Xiahai Zhuang$^*$\\
School of Data Science, Fudan University, Shanghai, China}

\begin{document}

\maketitle

\begin{abstract}
The principal rank-one (RO) components of an image represent the self-similarity of the image, which is an important property for image restoration.
However, the RO components of a corrupted image could be decimated by the procedure of image denoising.
We suggest that the RO property should be utilized and the decimation should be avoided in image restoration.
To achieve this, we propose a new framework comprised of two modules, i.e., the RO decomposition and RO reconstruction.
The RO decomposition is developed to decompose a corrupted image into the RO components and residual.
This is achieved by successively applying RO projections to the image or its residuals to extract the RO components.
The RO projections, based on neural networks, extract the closest RO component of an image.
The RO reconstruction is aimed to reconstruct the important information, respectively from the RO components and residual, as well as to restore the image from this reconstructed information.
Experimental results on four tasks, i.e., noise-free image super-resolution (SR), realistic image SR, gray-scale image denoising, and color image denoising, show that the method is effective and efficient for image restoration, and it delivers superior performance for realistic image SR and color image denoising.
\end{abstract}

\section{Introduction}\label{sec:introduction}
Image restoration aims at recovering an image from its degradation \cite{Milanfar/2013}.
The problem of image restoration is involved in many fields, such as low-level image processing \cite{Celebi/2014}, medical imaging \cite{Jan/2006}, and remote sensing \cite{Jensen/1995}. According to the process of degradation, it can be further categorized into more specific tasks, including image denoising, deblurring, and super-resolution (SR). 
Recently, the research on image restoration becomes particularly popular, thanks to the advance of deep learning and its introduction into this field.
Many methods have been proposed with promising performance \cite{Zhang_beyond/2017,Zhang_ffdnet/2018,Dong/2016,Ledig/2017}. Since fully automatic and real-time approaches are generally desirable in many applications, exploring such approaches becomes increasingly useful. 

Currently, the methods could be categorized into two groups, i.e., the model-based and the learning-based. The former models image restoration as a linear inverse problem \cite{Osher/2005,Elad/2006,Bioucas/2007,Candes/2009,Candes_matrix/2010}, and the latter learns a mapping from the degraded images to their ground truth \cite{Freeman/2002,Jain/2009,Timofte/2014,Schmidt/2014,Dong/2016}.
For the model-based methods, a widely used strategy is to solve the problem via the image priors driven regularizations, including local and non-local image priors. Local properties were widely used for image denoising, such as Tikhnove regularization \cite{Tikhnov/1977} and total variation (TV) regularization \cite{Rudin/1992}. Non-local self-similarity was extensively studied in many works, \textit{e.g.}, non-local means \cite{Buades/2005}, collaborative filtering \cite{Dabov/2007}, joint sparsity \cite{Mairal/2009}, and low-rank approximation \cite{Gu/2014}. For the low-rank approximation, matrices are often assumed to be low-rank.
Therefore, the methodologies of low-rank approximation \cite{Candes/2009,Gu/2014} are commonly used to solve the problem.
This strategy has two major advantages.
One is that an image can be well approximated by its low-rank components \cite{Andrews/1976,Candes/2009}, and the other is that the noise and low-rank components of an image can be easily separated from each other \cite{Candes_matrix/2010}. 
However, the methods of this group are also challenged by the complex parameter settings and expensive computation of the methods \textit{per se}. More specifically, one needs to manually set the initial hyper-parameters for the regularization terms, which can be application dependent; and as being iterative approaches, they need large numbers of iteration steps to ensure convergence.

\begin{figure}
\centering
\includegraphics[width=1\linewidth]{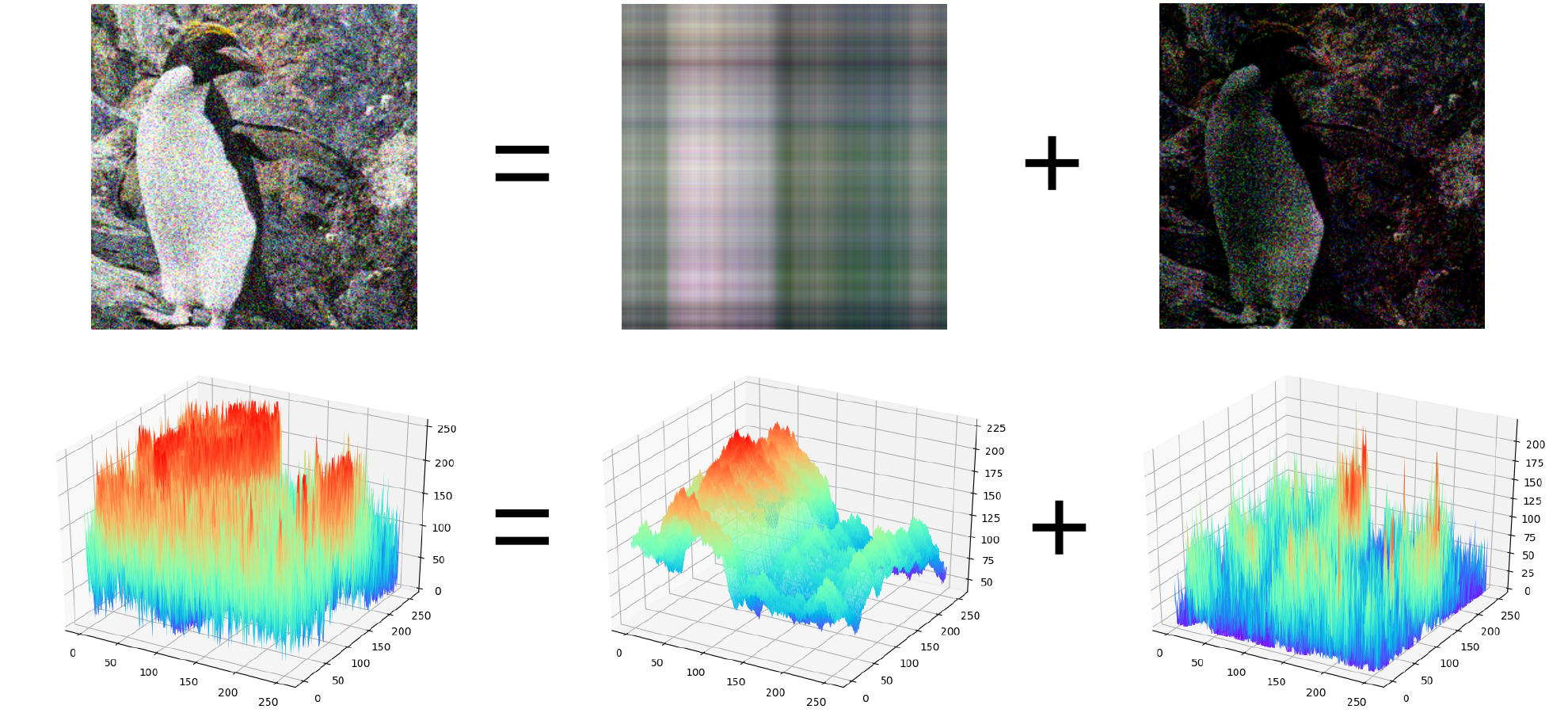}
\caption{Visualization of image decomposition obtained by our unsupervised RO decomposition. The top row shows that an noisy image could be decomposed into the sum of the almost noise-free low-rank part and the noisy residual error. We visualize the surfaces of their gray images at the bottom row, which shows that the low-rank component is more self-similar than the original image. One can refer to the text in Introduction section for details.}
\label{fig:00}
\end{figure}

The learning-based methods estimate a mapping from the degraded images to their ground truth, generally via dictionary \cite{Timofte/2014} or deep neural networks (DNNs) \cite{Kim/2016}. DNNs are powerful in modeling complex functions and thus have been widely explored for implementing the mapping \cite{Lempitsky/2018}.
A number of efficient techniques have been proposed to train the DNN models, such as the specific initializations \cite{Glorot/2010,He_delving/2015}, skip connection \cite{He/2016,Huang_densely/2017}, and batch normalization \cite{Loffe/2015}. Furthermore, the DNN methods can be easily implemented via well-developed parallel computing platforms, and the resulting models are computationally efficient during the testing stage. Particularly, the convolutional neural networks (CNNs) can effectively extract the high-order features of images, thus greatly improve the performance of learning-based methods in image restoration. Hence, they have been proposed for image denoising \cite{Jain/2009,Zhang_beyond/2017,Zhang_ffdnet/2018}, deblurring \cite{Chen/2016,Dong/2019}, and super-resolution \cite{Dong/2016,Kim/2016,Lim/2017}. For example, CNNs have been shown to achieve highly promising performance in removing Gaussian noise \cite{Chen/2016,Zhang_beyond/2017,Zhang_ffdnet/2018}; the methods combining deep residual neural networks deliver state-of-the-art performance in reconstructing the details of images \cite{Lim/2017,Wang/2018}.
Although the learning-based methods have been widely studied in the past, approaches of explicitly using non-local self-similarity with DNNs are rarely studied.
The property, however, has been shown to be crucial for image restoration \cite{Dabov/2007,Liu/2018}.
To rectify the weakness, a few pioneer works were reported to develop DNNs to learn the image prior and showed to be effective in improving the performance of image restoration \cite{Liu/2018,Lempitsky/2018}. Recently, neural nearest neighbors networks were proposed to achieve non-local processing by relaxing the traditional $ k $-nearest neighbors \cite{Ploetz/2018}, thus delivering state-of-the-art performance for image denoising. Nevertheless, there is no method to explicitly capture the low-rank property of images with DNNs, to the best of our knowledge.

In this work, we present the first attempt to employ the \emph{low-rank property} of degradations with DNNs.
The low-rank approximation of an image can be expressed as the sum of a series of rank-one (RO) components.
Furthermore, the RO components of an image are particularly self-similar, since any two rows or columns of an RO matrix are linearly dependent \cite{Andrews/1976}.
Therefore, we propose a framework based on RO decomposition and neural networks, referred to as rank-one network (RONet), to extract RO components and use them for the restoration of images.

RONet combines the idea of low-rank approximation with the learning-based scheme, to maintain their advantages. Concretely, an image corrupted by Gaussian noise could be decomposed into the sum of its low-rank component and residual error, as shown in Fig \ref{fig:00}, where the low-rank image is the sum of its first three RO components. Compared with the original image, the low-rank image is particularly self-similar and almost noise-free.
For efficient and effective implementation, we propose to extract RO components via a DNN specified for decomposition and utilize them to restore images via a DNN specified for reconstruction.
Specifically, we first decompose an image into the sum of RO components and residual error via a cascaded neural network. Then, the RO components and residual error are separately processed, since the former could be enhanced while the latter would be smoothed.
Finally, the recovered RO components and residual error are combined to approximate the ground truth.

The contributions of this work are summarized as follows.
\begin{itemize}
\item Firstly, we propose a new framework, based on RO decomposition and reconstruction, for image restoration.
The property of self-similarity of images exists in most of the real-world data.
However, this property has not been widely and explicitly considered for image restoration in the DNN-based approaches, to the best of our knowledge.
We develop this new framework, where the RO decomposition is aimed to exploit the self-similar features of an image, and the RO reconstruction is employed to recover the image from these features.
\item
Secondly, we implement the framework using DNNs, i.e., the RONet, which is comprised of the RO decomposition network and the RO reconstruction network. RONet is efficient in the test stage compared to non-learning-based algorithms.
\item
Finally, we verify our proposal that the RONet can separate the noises from the RO components of an image. This function has a particular advantage in realistic image restoration. We use four image restoration tasks to illustrate the performance of the proposed RONet.
\end{itemize}

The rest of our paper is organized as follows. In Section~\ref{sec02}, we introduce the related works about image restoration and RO decomposition.
Section~\ref{sec03} presents the methodologies of the proposed RONet.
Section~\ref{sec04} provides the implementation details of RONet and the evaluation results of four image restoration tasks. We finally conclude this work in Section~\ref{sec06}.

\begin{figure*}[tb]
\centering
\includegraphics[width=1\linewidth]{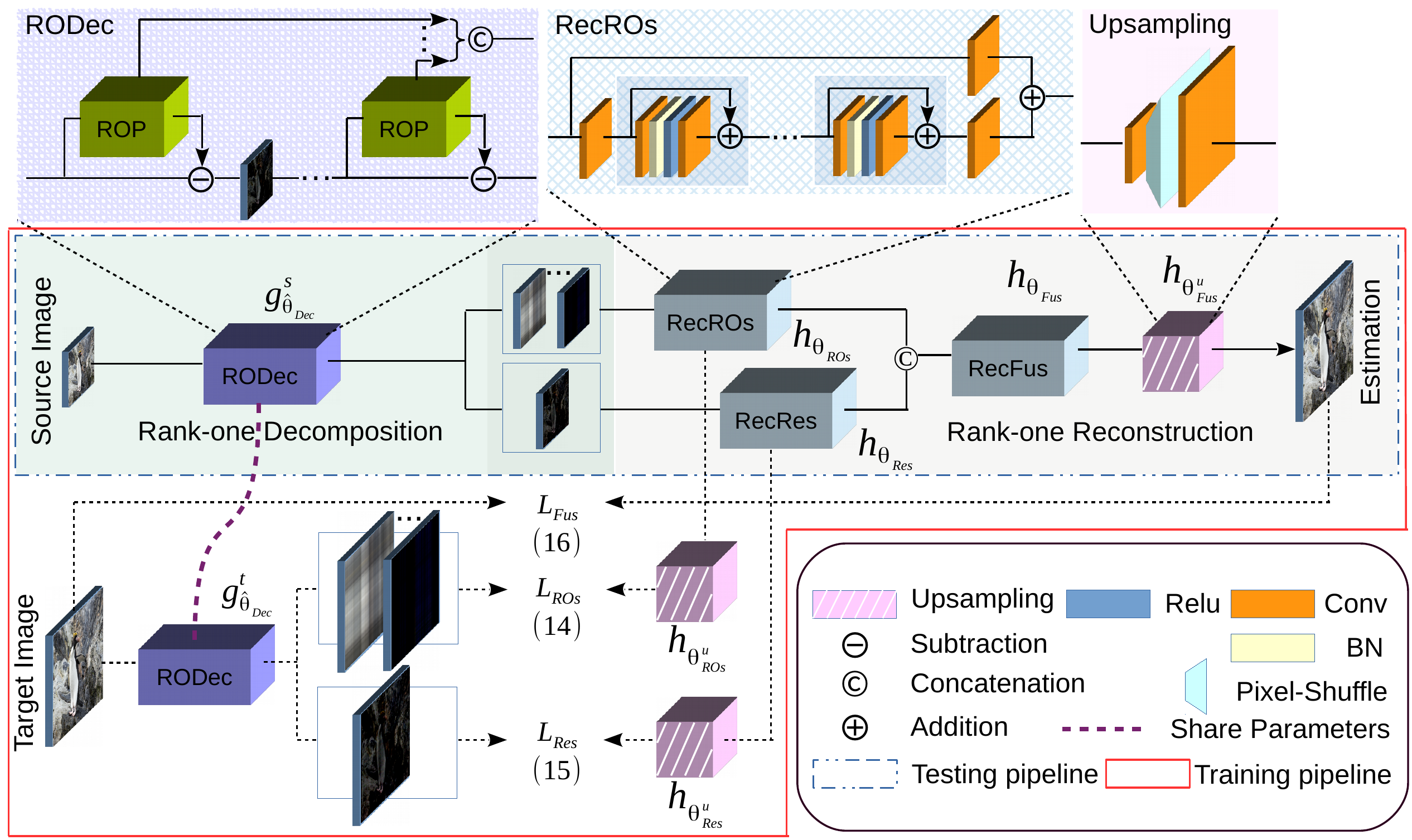}
\caption{Pipelines of the proposed method. Here, \textit{RODec} denotes the main body of our RO decomposition network, which is pre-trained by the loss $ L^{unsup}_{Dec} $ in (\ref{eq16}), \textit{RecRos}, \textit{RecRes}, \textit{RecFus}, and \textit{Upsampling} (only for image super-resolution) consist of the main body of our RO reconstruction network, which is trained by the loss $ L_{Rec} $ in (\ref{eq15}). \textit{RecROs}, \textit{RecRes}, and \textit{RecFus} are aimed at reconstructing RO components, restoring residual, and fusing the concatenation, respectively. They share the similar architecture but have different depths.
	ROP network is illustrated in Fig \ref{fig:02}.}
\label{fig:01}
\end{figure*}

\section{Related work}\label{sec02}
\subsection{Image restoration}
Recovering images from a few corrupted observations is particularly difficult, which appears in single image super-resolution (SISR) and image denoising. Conventionally, one could model this task as a linear inverse problem, and solve it using the approaches of regularization \cite{Rudin/1992,Osher/2005,Shi/2015}. With the success of deep learning, many deep neural network (DNN)-based approaches have been developed for image restoration with promising progress \cite{Jain/2009,Sun/2015,Dong/2016}.

SISR is the task of reconstructing the high-resolution images from the low-resolution images. It is very difficult when the upscaling factor is more than 4 and when low-resolution images are corrupted with unknown noise.
Interpolation is often used for SISR, but it obtains poor results due to its effect on smoothing details. Recently, DNNs have set state of the art for SISR. As a pioneer work, Dong et al. \cite{Dong/2016} developed a three layers network, referred to as SRCNN, to learn the end-to-end mapping from the low-resolution to the high-resolution images. Inspired by the superior performance of SRCNN, many DNNs have been proposed to improve the Peak Signal-to-Noise Ratio (PSNR) value, such as the residual learning based networks \cite{Lim/2017,Sajjadi/2017,Kim1/2016,Yu/2018}, the recursive learning based networks \cite{Kim/2016}, the dense connection-based networks \cite{Zhang/2018,Tong/2017}, and the Laplacian pyramid based networks \cite{Lai/2017}. Besides, to enhance the visual quality of SR images, the perceptual loss \cite{Johnson/2016} and the adversarial learning based networks \cite{Ledig/2017,Wang/2018} have been proposed. But these approaches would lower the PSNR value, compared to the PSNR-oriented approaches. Therefore, achieving a good trade-off between the PSNR value and the visual quality of SR is still challenging.

Image denoising is an essential task in computer vision since data is inevitably accompanied by noise during the imaging process. Conventional approaches mainly concentrate on modeling the image priors explicitly, such as the low-rank based methods \cite{Gu/2014} and the total variation based methods \cite{Rudin/1992}. Dabov et al. \cite{Dabov/2007} proposed a competitive method, referred to as BM3D, for image denoising due to its flexibility in dealing with the various noise levels. Recently, deep CNN-based approaches have made great progress in image denoising. Burger et al. \cite{Burger/2012} proposed a multi-layer perceptron method that achieves comparable performance with BM3D. Inspired by the promising performance of deep CNN, Zhang et al. \cite{Zhang_beyond/2017} developed a residual learning-based network, which delivered superior denoising performance against the previous works. Besides, Zhang et al. \cite{Zhang_ffdnet/2018} proposed a fast CNN-based network to increase the flexibility of dealing with the spatially variant noises. However, achieving a good trade-off between the inference speed and the denoising performance is still challenging.
\subsection{RO decomposition}
Rank-one decomposition aims at factorizing a matrix into the sum of rank-one components. One of the well-known methods is the singular value decomposition (SVD). SVD is the optimal rank-one decomposition in the least square sense, and it becomes a basic tool in matrix analysis thanks to its powerful ability in extracting the low-rank components of a matrix. Exploring the rank-one decomposition is particularly attractive since it is the basis of matrix completion \cite{Candes/2009,Hu_fast/2012} and principle component analysis (PCA) \cite{Candes/2011}.

Both matrix completion and PCA show that extracting the low-rank components is significant since the components are compressed and have high self-similarity. In computer vision, image restoration refers to study the self-similarity of an image. Therefore, we could use the techniques of matrix completion and PCA to estimate the low-rank components of an image. Since the low-rank matrix is often expressed as the sum of RO matrices, it is significant to explore RO decomposition and its application.

Matrix completion is involved in many areas, including machine learning \cite{Amit/2007,Argyriou/2007}, computer vision \cite{Komodakis/2006}, and recommender system \cite{Koren/2008}. Recent progress in matrix completion shows that low-rank components are good approximations of unknown matrices. Under some general constraints, one can perfectly recover low-rank components from incomplete observations with very high probability \cite{Candes/2009}. Since the generic denoising models assume that the noise can be separated from data, low-rank components could be reconstructed accurately even if the observed entries are corrupted with a small number of noises \cite{Candes_matrix/2010}.

Classical PCA shows that the low-rank components of data are its principal components \cite{Hotelling/1933,Jolliffe/1986}. Recently, splitting the sparse components becomes another growing interest, and several works have been done to decompose a matrix into the sum of the low-rank and sparse components. Chandrasekaran et al. \cite{Chandrasekaran/2011} showed the conditions of exactly recovering the low-rank and sparse components by developing a notion of rank-sparsity incoherence. Robust PCA proposed a convenient method to recover both the low-rank and sparse components, and presented its wide applications in video surveillance and face recognition \cite{Candes/2011}. Besides, tensor robust PCA was proposed to decompose a tensor into the sum of low-rank and sparse tensors, and it performs robustly in image restoration and background modeling \cite{Lu_tensor/2019}.

\begin{table}[t]
\centering
\caption{Summarization of the notions and notations used in the methodology.}
\label{tab01}
\begin{tabular}{|c|c|}
	\hline
	Notions&  Notations\\
	\hline
	Image/RO space&  $ \mathbb R^{m\times n} $/$ O^{m\times n} $\\
	\hline
	Number of samples/ROPs & $ N $/$ L $\\
	\hline
	$ L $-level residual space&  $ R_L^{m\times n} $ \\
	\hline
	Cartesian space&  $ (O^{m\times n})^L\times R_L^{m\times n} $\\
	\hline
	Source/Target space&    $ \mathbb R^{m_s\times n_s} $/$ \mathbb R^{m_t\times n_t} $\\
	\hline
	Image/Residual& $ X $/$ E $\\
	\hline
	Source/Target image & $ S $/$ T $\\
	\hline
	Sample/ROP index  & Superscript/Subscript of $ X_l^i $\\
	\hline	
\end{tabular}
\end{table}

\section{Methodology}\label{sec03}

This work is aimed to develop an effective framework for image restoration.
Particularly, image restoration from noisy observations can be challenging, since the progress of denoising in image space could corrupt their details.
Therefore, we propose a two-stage framework for image restoration, i.e., (1) decomposition of the source image into two parts, particularly into a series of RO components and the residual error, and (2) reconstruction of the target image from the RO components and the residual error. This framework is effective in recovering the corrupted details since the RO components are often noise-free. SVD and the techniques of matrix completion can be applied to perform the decomposition and reconstruction, respectively. However, they are the iterative methods and are computationally expensive when the sizes of images are large. Inspired by the powerful ability of deep neural networks in learning complex mappings, we propose to obtain the mappings of source image decomposition and target image reconstruction via deep learning.

Fig \ref{fig:01} illustrates the pipelines and network architectures of the proposed method.
\textbf{In the decomposition stage}, we build a cascaded CNN, referred to as \textit{RODec}, using the proposed cascade unit (RO projection).
This unit projects a matrix into two parts, i.e., an RO matrix and a residual error matrix.
We implement this projection using a DNN and refer to this network as RO projection network (ROPNet) for convenience in Section \ref{sec3.1}.
As a result, we can decompose the source image into the sum of a series of RO components and a residual error using the RO projections.
\textbf{In the reconstruction stage}, we build three residual CNNs, referred to as \textit{RecROs}, \textit{RecRes}, and \textit{RecFus}
for the reconstruction of target images and estimation of residual errors as Fig \ref{fig:01} shows.
The \textit{Upsampling} block is to ensure the low dimension of operating space.

Table \ref{tab01} summarizes the important notations used in this paper, where $ \mathbb R^{m\times n} $ denotes image space, and $ O^{m\times n} $ denotes the corresponding RO space. In the manuscript, we use the subscripts $ s $ and $ t $ to denote the notations related to the source and target image spaces, respectively. Note that $ \left\| X \right\|_2  $ denotes $ \ell_2 $ norm of the vectorization of tensor $ X $, and $ \left\| X \right\|_1  $ denotes $ \ell_1 $ norm of the vectorization of $ X $, unless stated otherwise. The rest of this section is organized as follows.
First, we introduce the RO projection in Section \ref{sec3.1}, which is the core of RO decomposition.
Then, we elaborate on the decomposition and reconstruction in Section \ref{sec3.2} and Section \ref{sec3.3}, respectively.
Finally, we describe the details of the training and testing of the networks in Section \ref{sec3.4}

\subsection{RO projection}\label{sec3.1}

\begin{figure}[!t]
\centering
\includegraphics[width=0.8\linewidth]{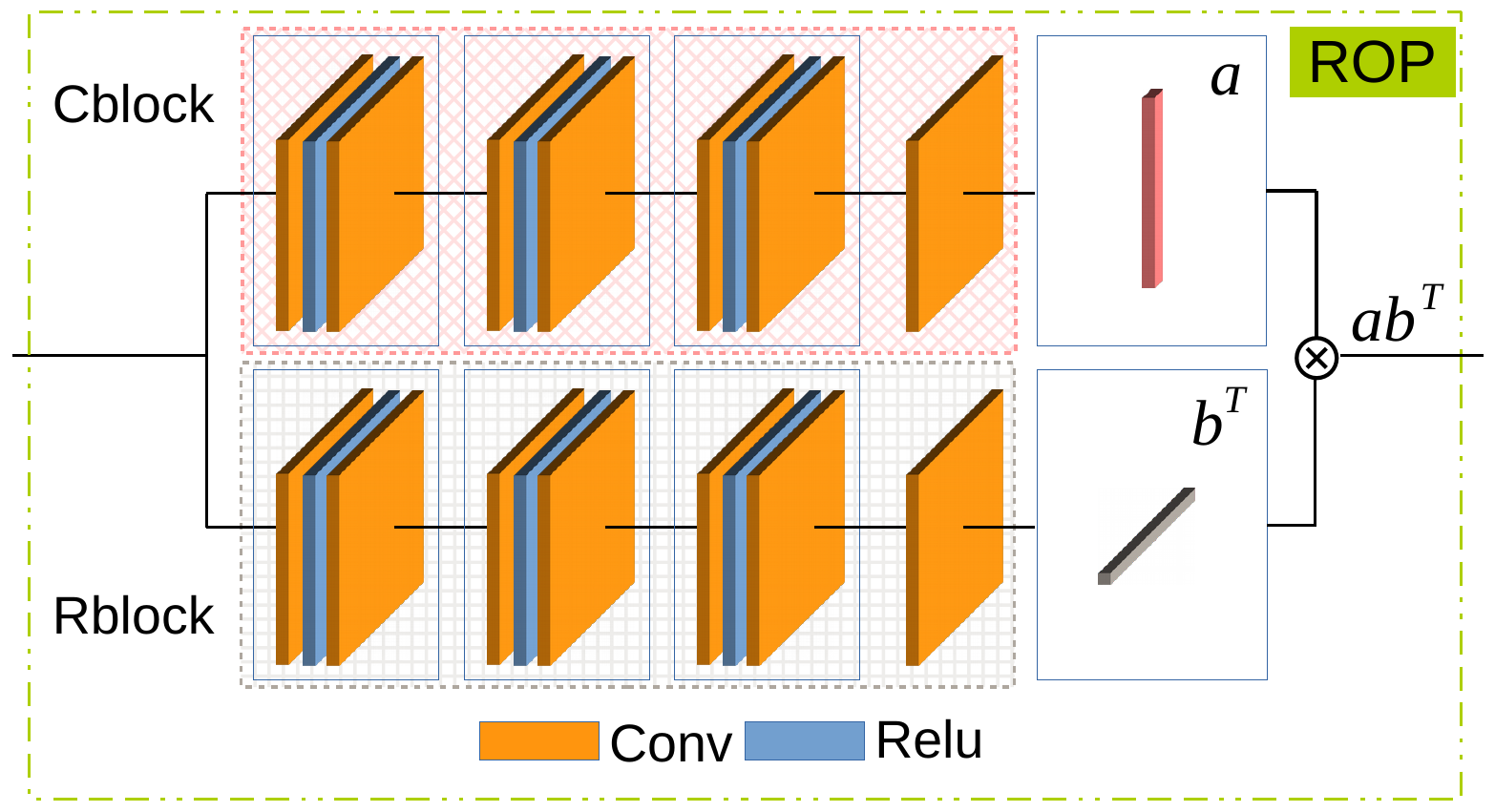}
\caption{The architecture of the rank-one projection network (ROPNet), which is aimed at projecting the input to be an RO matrix.}
\label{fig:02}
\end{figure}

\emph{A matrix is rank-one if and only if it can be expressed as the outer product of two vectors}, which leads to the advantages that RO matrices are highly self-similar and can be greatly compressed.
Real-world images are matrices in image spaces, whose ranks are generally much greater than one due to their complex contents.
The simple RO components could be used to represent the principal part of an image \cite{Hotelling/1933,Jolliffe/1986}.
Therefore, we propose to employ RO projections to obtain the RO components of an image for image restoration.

Let $ \mathbb R^{m \times n } $ be an image space, and $ \lbrace X^i \rbrace_{i=1}^N  $ denote $ N $ samples from $ \mathbb R^{m \times n } $. Suppose $ \Theta_p $ is a parameter space, then an RO projection network (ROPNet), denoted as $ f_{\theta_p} $ with parameters $ \theta_p\in \Theta_p $, is a mapping from $ \mathbb R^{m \times n } $ to $ O^{m \times n } $, i.e., $ f_{\theta_p} : \mathbb R^{m\times n} \rightarrow O^{m\times n} $, where $ O^{m \times n } = \lbrace X\in \mathbb R^{m\times n} | \mbox{rank}(X) = 1 \rbrace  $ denotes the RO space with respect to $ \mathbb R^{m\times n} $. Therefore, the RO projection space generated by the ROPNet can be expressed as $ F = \lbrace f_{\theta_p} | \theta_p\in \Theta_p \rbrace  $. Our aim is to obtain proper parameters such that
\begin{equation}\label{eq01}
\hat{\theta}_p \in \mathop{\arg\min}_{\theta_p \in \Theta_p} \frac{1}{N} \textstyle\sum_{i=1}^{N}\left\| X^i - f_{\theta_p}(X^i)\right\|_2^2.
\end{equation}

We propose to employ CNNs to build the ROPNet and learn the solution of problem (\ref{eq01}). The ROPNet is aimed to learn proper RO projections from the image space to the RO space. For each RO matrix, it could be expressed as the matrix product of a column vector and a row vector.
Therefore, we design the ROPNet that consists of two branches.
One aims at projecting a given matrix to a column vector, and the other maps the matrix to a row vector.

As Fig \ref{fig:02} shows, each branch of the ROPNet consists of two functional mappings, i.e., a non-linear mapping and vectorization mapping.
The former is comprised of a number of convolutional layers and non-linear activation layers;
the latter is designed as the average pooling in a row or column direction, which is aimed at projecting a matrix to a column or row vector. The mathematical motivation is given in Appendix A of Appendices.
Therefore, ROPNet projects a color image into three RO matrices, of which each is the matrix product of the corresponding column vector and row vector projected by the vectorization mapping.

\subsection{RO decomposition}\label{sec3.2}
RO decomposition maps the image space to the Cartesian product of a set of RO spaces and a residual space, i.e., decomposes an image into a series of RO matrices and a residual error.
For each $ X \in \mathbb R^{m\times n} $, its RO decomposition could be expressed as follows,
\begin{equation}\label{eq02}
X = X_1 + X_2 + \cdots + X_L + E_L,
\end{equation}
where, $ X_l \in O^{m\times n} $ ($ 1\le l \le L $) denotes the $ l $-th RO matrix; $ E_L \in \mathbb R^{m\times n} $ denotes the $ L $-th residual error; and $ L $ denotes the number of RO components, which is user-defined. Such decomposition as (\ref{eq02}) shows is known to be unique and determined by SVD in the sense of least squares, namely, the unique decomposition could be obtained by solving the following problem,
\begin{equation}\label{eq03}
\{\hat{X}_i\} = \mathop{\arg\min}_{\left\lbrace X_i\in O^{m\times n}\right\rbrace }\frac{1}{2}|| X - \textstyle \sum_{i=1}^L X_i||_2^2.
\end{equation}
Moreover, the first few RO components of an image are generally free from noises \cite{Andrews/1976,Candes_matrix/2010},
which motivates us to extract these noise-free components of a given image for restoration.

To obtain an appropriate RO decomposition as SVD, we solve a series of RO estimation problems. Concretely, an RO decomposition could be achieved by successively applying a series of ROPNets, i.e., $ \lbrace f_{\theta_p^l} \rbrace_{l=1}^L  $, to an image and the residuals after each ROPNet. For the $ i $-th image sample $ X^i \in \mathbb R^{m\times n} $, we could obtain a series of RO components and the corresponding residual via the following formulas,
\begin{gather}
X_1^i = f_{\theta_p^1}(X^i),\label{eq031}\\
E_l^i = X^i - \textstyle\sum_{j=1}^l X_j^i,\label{eq032}\\
X_{l+1}^i = f_{\theta_p^{l+1}}( E_l^i),\label{eq033} 
\end{gather}
for $ l = 1, \dots, L-1 $.
Here, $ X_1^i $ and $ E_1^i $ depend on $ \lbrace \theta_p^1 \rbrace  $, $ X_{l+1}^i $ and $ E_{l+1}^i $ depend on $ \lbrace \theta_p^1, \ldots, \theta_p^{l+1} \rbrace $. For convenience, we introduce a new notation $ R^{m\times n}_l $, which is referred to as the $ l $-level residual space, i.e., $ R^{m\times n}_l=\lbrace E_l| X \in \mathbb R^{m\times n}; \theta_p^j \in \Theta_p, \forall j\in \lbrace 1, \dots, l\rbrace \rbrace $, where $ E_l $ is determined by $ X $ and $ \lbrace \theta_p^j \rbrace_{j=1}^{l} $ as (\ref{eq031})-(\ref{eq033}) illustrate. To obtain an appropriate decomposition, we use the strategy as (\ref{eq01}) shows to learn the parameters of ROPNets in the sense of least squares. Concretely, the proper parameters are obtained by successively minimizing the residual errors from the $ 1 $- to $ L $-level residual space, as follows,   
\begin{equation}\label{eq034}
\hat{\theta}_p^l \in \mathop{\arg\min}_{\theta_p^l \in \Theta_p} \frac{1}{N}\textstyle\sum_{i=1}^{N} \| E_{l-1}^i - f_{\theta_p^l}(E_{l-1}^i) \|_2^2,
\end{equation}
where, $ l=1, \dots, L $ and $E_{0}^i=X^i$. Note that $ \hat{\theta}_p^l $ depends on $ \lbrace E_{l-1}^i \rbrace_{i=1}^N $, and the precise formula of $ \hat{\theta}_p^l $ is $ \hat{\theta}_p^l(E_{l-1}^1, \dots, E_{l-1}^N) $. Here, we omit the dependence to simplify notation.

Let the mappings from the image space to the Cartesian space be denoted as the elements in
\begin{equation}\label{eq035}
G =\lbrace g_{\theta_{Dec}} | g_{\theta_{Dec}} : \mathbb R^{m\times n} \rightarrow (O^{m\times n})^L\times R_L^{m\times n} \rbrace
\end{equation}
such that $ g_{\theta_{Dec}}(X^i) = (X_1^i, \dots, X_L^i, E_L^i) $, where $ (O^{m\times n})^L $ denotes the $ L $-Cartesian product of RO spaces, and $ \theta_{Dec} = \lbrace \theta_p^l \rbrace_{l=1}^L $ is in $\Theta_p^L $, which denotes the $ L $-Cartesian product of parameter spaces. The formulas in (\ref{eq034}) and (\ref{eq032}) induce that we could obtain the optimal parameters $ \lbrace \hat{\theta}_p^l\rbrace_{l=1}^L  $ by successively minimizing the residual, i.e., $ E_l^i = E_{l-1}^i - X_l^i $, from the low-level to the high-level residual space. However, we jointly optimize the parameters in practice. That is we obtain the optimal parameters for RO decomposition by minimizing the following problem,
\begin{align}\small
\hat{\theta}_{Dec} &\in \mathop{\arg\min}_{\theta_{Dec}\in \Theta_p^L} \frac{1}{NL}\textstyle\sum_{i=1}^N\textstyle\sum_{l=1}^L \| E_{l-1}^i - X_l^i \|_2^2\label{eq04}\\  &\in  \mathop{\arg\min}_{\theta_{Dec}\in \Theta_p^L} \frac{1}{NL} \textstyle\sum_{i=1}^N\sum_{l=1}^L \| X^i - \sum_{j=1}^l [ g_{\theta_{Dec}}(X^i) ] _j \|_2^2,\nonumber
\end{align}
where, $ [ g_{\theta_p}(X^i) ] _j $ denotes the $ j $-th element of $ g_{\theta_p}(X^i) $.

The formulas (\ref{eq031})-(\ref{eq033}) show that ROPNets could be applied to obtain the RO decomposition of an image.
Therefore, we utilize the ROP as the basic unit to build the RO decomposition network (RODec). Note that once a network RODec has been successfully trained, it can be used in different applications where one needs to decompose an image into the sum of RO components. For example, the RODec trained using the dataset DIV2K \cite{ntire/2018} was used for the tasks of image super-resolution and image denoising in our experiments. For the specific image restoration, we introduce two extra image spaces to avoid confusion, i.e., source space and target space, which are notated as $ \mathbb R^{m_s\times n_s} $ and $ \mathbb R^{m_t\times n_t} $, respectively. There are two RO spaces with respect to them, i.e., source RO space and target RO space, which are notated as $ O^{m_s\times n_s} $ and $ O^{m_t\times n_t} $, respectively. Besides, we use $ \lbrace S^i, T^i \rbrace_{i=1}^N  $ to denote $ N $ paired samples from the source and target space.

The architecture of RODec is shown at the upper left corner of Fig \ref{fig:01}.
Concretely, for the $ i $-th source image $ S^i \in \mathbb R^{m_s\times n_s}$, we still denote it as $ E_0^i $ for convenience. First we use the ROP parameterized with $ \theta_p^1 $ to project this image to an element $ S_1^i \in O^{m_s\times n_s} $, and compute the first residual $ E_1^i $ in the $ 1 $-level residual space $ R_1^{m_s\times n_s}  $, i.e., $ S_1^i = f_{\theta_p^1}(E_0^i) $ and $ E_1^i = E_0^i - S_1^i $. Then we utilize the ROP parameterized with $ \theta_p^2 $ to project $ E_1^i $ to the second RO component $ S_2^i $, and calculate the residual $ E_2^i \in R_2^{m_s\times n_s} $, i.e., $ S_2^i = f_{\theta_p^2}(E_1^i) $ and $ E_2^i = E_1^i - S_2^i $.
Similarly, we project the $ (L-1) $-th residual error to an element $ S_L^i \in O^{m_s\times n_s} $, and obtain the residual $ E_L^i $ in the $ L $-level residual space, i.e., $ S_L^i = f_{\theta_p^L}(E_{L-1}^i) $ and $ E_L^i = E_{L-1}^i - S_L^i $.
Finally, the outputs of \textit{RODec} are composed of the residual error $ E_L^i $ and the series of $ S_1^i$,$ \dots $, and $ S_L^i $.

To train RODec, we could use the unsupervised method as shown in (\ref{eq04}). Besides, we can also use supervised learning by minimizing the following loss function,
\begin{align}\small
L_{Dec}^{sup} &=\frac{1}{NL}\textstyle\sum_{i=1}^{N}\textstyle\sum_{l=1}^L\| S_l^i - \tilde{S}_l^i  \|_2^2\nonumber\\
&= \frac{1}{NL}\textstyle\sum_{i=1}^{N}\textstyle\sum_{l=1}^L \| \left[ g_{\theta_{Dec}}(S^i)\right]_l - \tilde{S}_l^i \|_2^2, \label{eq09}
\end{align}
where, $ \lbrace \tilde{S}_l^i \rbrace_{l=1}^L  $ denotes the least square solution of (\ref{eq03}), which is calculated by SVD, for the $ i $-th source image $ S^i $.

Although SVD can be used to obtain the RO decomposition in the least square
sense, it is computationally expensive for the proposed learning-based method, which repeatedly applies the RO decomposition for each image in the training stage. That will be shown in our ablation study. Note that the RO components obtained by RODec are not the same as the results computed by SVD, since SVD is calculated for a particular image while RODec trained on a set of samples.

\subsection{RO reconstruction}\label{sec3.3}
We build the mappings from the Cartesian space to the target space, so as to reconstruct the target images. Most of the learning-based approaches have been developed directly from the source space to the target space without explicitly considering the self-similarity of degradations \cite{Jain/2009,Dong/2016,Chen/2016}. However, the previous works have shown that utilizing self-similarity could improve the performance and robustness of learning-based approaches \cite{Dabov/2007,Lempitsky/2018,Liu/2018}. Since RODec has the ability to extract the highly self-similar RO components, we develop a network from the source Cartesian space to the target space to sufficiently utilize the RO components.

Image restoration is often modeled as a linear problem. Concretely, let $ x^i\in \mathbb R^n $ be the vectorization of a target image $ T^i\in \mathbb R^{m_t\times n_t} $, and $ y^i\in \mathbb R^m $ be the vectorization of a source image $ S^i\in \mathbb R^{m_s\times n_s} $, where $ n = m_t n_t $, $ m = m_s n_s $, and $ \frac{m_t}{m_s} $ and $ \frac{n_t}{n_s}$ are upscaling factors.
Suppose $ A\in \mathbb R^{m\times n} $ is a degradation operator, then the image restoration problem could be expressed as $ y^i = Ax^i + \epsilon $, where $ \epsilon $ denotes  noise. For the task of image denoising, the upscaling factor is set to be $ 1 $, and $ A $ is an identity matrix. For the task of image super-resolution, the upscaling factors are generally set to be integers more than 1, and $ A $ is a downsampling matrix.

Many learning-based methods aim at estimating a mapping directly from the source space to the target space. In contrast, we build a mapping from source Cartesian space to the target space due to the ability of RODec in extracting RO components. Let $ (O^{m_s\times n_s})^L \times R_L^{m_s\times n_s} $ denote the source Cartesian space,  $ (O^{m_t\times n_t})^L \times R_L^{m_t\times n_t} $ denote the target Cartesian space, and $ \Theta_r $ denote a parameter space for reconstruction. The mapping from the source Cartesian space to the target space is defined by $ h_{\theta_{Rec}} $, which is parameterized by $\theta_{Rec}$, and $\theta_{Rec}\in \Theta_r $. Given a training dataset $ \lbrace S^i, T^i \rbrace_{i=1}^N  $, we then want to find the proper parameters by the following minimization,
\begin{equation}\small\label{eq10}
\hat{\theta}_{Rec}\in \mathop{\arg\min}_{\theta_{Rec}\in \Theta_r} \frac{1}{N}\textstyle\sum_{i=1}^N \| h_{\theta_{Rec}}(\hat{S}_1^i, \dots, \hat{S}_L^i, \hat{E}_L^i) - T^i \|_2^2,
\end{equation}
where, $ (\hat{S}_1^i, \dots, \hat{S}_L^i, \hat{E}_L^i) = g_{\hat{\theta}_{Dec}}^s(S^i) $, and $ g_{\hat{\theta}_{Dec}}^s(\cdot)$ denotes a pre-trained RODec. The optimal solution, however, is hard to obtain due to the fact that there are a large number of parameters to optimize for DNNs. Moreover, we would get a local minimum which may be greatly deviated from the optimal solution. We therefore adopt other priors to construct regularization losses, so as to obtain a more robust solution.

Taking single image super-resolution (upscaling factor $ \times 4 $) as an example, we utilize residual blocks to design an RO reconstruction network (RORec) for learning the mappings from the source Cartesian space to the target space. As shown in Fig \ref{fig:01}, we use \textit{RecROs}, \textit{RecRes}, \textit{RecFus}, and \textit{Upsampling} to build our RORec.
\textit{RecROs} is built to reconstruct the RO components of a target image, meanwhile, \textit{RecRes} is proposed to estimate the residual error of the target image. \textit{RecFus} is developed to obtain the fusion of the RO components and residual error, which is taken as an estimation of the target image. Note that the three blocks share the similar network structure as shown in Fig \ref{fig:01}, but have different depths. \textit{Upsampling} is built to ensure that most of the operators are implemented in a low-dimensional space. \textit{Upsampling} with respect to \textit{RecFus} consists of convolutional layers and pixel-shuffle. Note that for image denoising  this upsampling is unnecessary and thus is removed. For convenience, we denote \textit{RecROs}, \textit{RecRes}, \textit{RecFus}, and \textit{Upsampling} of RORec as $ h_{\theta_{ROs}} $, $ h_{\theta_{Res}} $, $ h_{\theta_{Fus}} $, and $ h_{\theta_{Fus}^u} $, respectively, as shown in Fig \ref{fig:01}. Then, we have $ \theta_{Rec} = \lbrace \theta_{ROs},\theta_{Res},\theta_{Fus}, \theta_{Fus}^u\rbrace $. For training RORec, we are going to use the RO decomposition of target images. Therefore, extra two \textit{Upsampling} blocks with respect to \textit{RecROs} and \textit{RecRes} are included, which are denoted as $ h_{\theta_{ROs}^u} $ and $ h_{\theta_{Res}^u} $, respectively. Overall, the mappings in Fig \ref{fig:01} are summarized in Appendix B of Appendices.

We not only compute a loss in the target space, but also design two losses in the target Cartesian space. Concretely, given a source image $ S^i\in \mathbb R^{m_s\times n_s} $,  we could use the pre-trained RODec $ g_{\hat{\theta}_{Dec}}^s $ to map the image to its RO components and residual error, i.e., we have
\begin{align}\label{eq11}
\hat{S}^i_{ROs}&=( [ g_{\hat{\theta}_{Dec}}^s(S^i)]_1, \dots, [ g_{\hat{\theta}_{Dec}}^s(S^i)]_L ) ,\\
\hat{S}^i_{Res}&=[ g_{\hat{\theta}_{Dec}}^s(S^i)]_{L+1}.
\end{align}
Moreover, we utilize $ g_{\hat{\theta}_{Dec}}^t $, which shares the same parameters with $ g_{\hat{\theta}_{Dec}}^s $, to obtain the low-rank component and residual error of the target $ T^i\in \mathbb R^{m_t\times n_t} $, namely,
\begin{equation}\small\label{eq11a}
\hat{T}^i_{LR}=\textstyle\sum_{l=1}^L [ g_{\hat{\theta}_{Dec}}^t(T^i)]_l \mbox{ and }
\hat{T}^i_{Res}=[ g_{\hat{\theta}_{Dec}}^t(T^i)]_{L+1},
\end{equation}
where, $ [ g_{\hat{\theta}_{Dec}}^t(T^i)]_l\in O^{m_t\times n_t}  $ for $ l = 1, \dots, L $, and $ [ g_{\hat{\theta}_{Dec}}^t(T^i)]_{L+1}\in R_L^{m_t\times n_t} $. Then the loss functions for RO components and residual error are given by
\begin{align}
L_{ROs} &= \frac{1}{N} \textstyle\sum_{i=1}^{N} \| h_{\theta_{ROs}^u}\circ h_{\theta_{ROs}}(\hat{S}^i_{ROs})  -  \hat{T}^i_{LR} \|_{\alpha}^{\alpha} ,\\
L_{Res} &= \frac{1}{N} \textstyle\sum_{i=1}^{N} \| h_{\theta_{Res}^u}\circ  h_{\theta_{Res}}(\hat{S}^i_{Res}) - \hat{T}^i_{Res} \|_{\alpha}^{\alpha}.
\end{align}
where, $ \circ $ denotes the composition of two functions.
The loss for fusion is expressed as
\begin{align}
L_{Fus} =& \frac{1}{N} \textstyle\sum_{i=1}^{N} \| h_{\theta_{Rec}}\circ g_{\hat \theta_{Dec}}^s (S^i) - T^i \|_{\alpha}^{\alpha} + \nonumber\\
& \eta L_{Per}\left( h_{\theta_{Rec}}\circ g_{\hat \theta_{Dec}}^s (S^i), T^i\right)  ,
\end{align}
where, $ \alpha\in \left\lbrace 1, 2\right\rbrace  $, $ L_{Per} $ denotes the perceptual loss \cite{Johnson/2016}, and $ \eta $ is a hyper-parameter. Overall, our RORec is trained by minimizing the following loss function,
\begin{equation}\label{eq15}
L_{Rec} = \lambda (L_{ROs} + L_{Res}) + (1-\lambda) L_{Fus},
\end{equation}
where, $ \lambda\in [0, 1] $ is another hyper-parameter. Note that we could also jointly train RODec and RORec without fixing the parameters of RODec for a particular task. In this case, the flexibilities of RODec in adapting different tasks would be lost, since RODec depends on a specific dataset conducted by the task.

\subsection{Training and testing}\label{sec3.4}
In the training phase, the RO decomposition network can be trained via both supervised and unsupervised learning. For the former, we minimize the distance between the RO components and the ground truth obtained by \textit{RODec} and SVD, respectively. For the latter, we directly minimize the $ \ell_2 $ norm of the residuals induced by \textit{ROPs}. To train the RO reconstruction network, we develop three loss functions, i.e., $ L_{ROs} $, $ L_{Res} $, and $ L_{Fus} $. Here, $ L_{ROs} $ is the distance between the upsampling output of \textit{RecROs} and the low-rank component of the target image. $ L_{Res} $ denotes the distance between the upsampling output of \textit{RecRes} and the residual error of the target image. $ L_{Fus} $ represents the distance between the upsampling output of \textit{RecFus}  and the target image.
Overall, the RO reconstruction network can be trained by minimizing the weighted sum of the three losses as shown in (\ref{eq15}).

As shown in Fig \ref{fig:01}, RO decomposition networks $ g^s_{\hat{\theta}_{Dec}} $ and $ g^t_{\hat{\theta}_{Dec}} $ (sharing the same parameters) have been pre-trained before training the RO reconstruction network. Therefore, we train the RO decomposition network via both supervised and unsupervised methods firstly. For supervised learning, we could train RODec by minimizing the loss function in (\ref{eq09}) to get the pre-trained models. Let the objective function in (\ref{eq04}) denote as
{\small 
\begin{equation}\label{eq16}
L_{Dec}^{unsup} = \frac{1}{NL}\sum_{i=1}^N\sum_{l=1}^L \| X^i - \sum_{j=1}^l \left[ g_{\theta_{Dec}}(X^i) \right] _j \|_2^2.
\end{equation}}
Then we could minimize this loss function for unsupervised learning.

After obtaining the pre-trained RODec, we would fix its parameters and use supervised learning to train our RORec. In practice, the loss function varies from one task to another in image restoration. For single image super-resolution, we minimize the loss in (\ref{eq15}) by empirically setting $ \lambda = 0.5 $, $ \eta = 1\times 10^{-3} $, and $ \alpha=1 $. For image denoising, $ \lambda $ and $ \eta $ are both set to be $ 0 $, and $ \alpha $ is set to be 2, i.e., we use mean squared error to train RORec.

In the testing phase, a source image is first decomposed by \textit{RODec} into some RO components and a residual error. Then the outputs of \textit{RODec} are mapped by \textit{RecROs} and \textit{RecRes} to approximate the RO components and the residual error of the target image, respectively. Finally, the concatenation of previous outputs is mapped successively by the \textit{RecFus} and \textit{Upsampling} to be an estimation.

\section{Experiments}\label{sec04}

In the experiments, we first evaluated the performance of RODec in Section~\ref{sec:RODec}. Then, we performed the ablation studies to obtain the proper architectures and settings of the decomposition and reconstruction networks for the RONet, in Section~\ref{sec43}. Finally, we evaluated the performance of RONet for four image restoration tasks, in Section~\ref{exp:recon}, where the assumption that the process of denoising could corrupt the RO components of a noisy image is verified.

Note that for noise-free image restoration, the RONet framework may not perform particularly better than the other state-of-the-art methods, since RONet could not demonstrate its advantage in separating the RO components from the noise when no noise is presented in the images.
To illustrate this, we evaluated a RONet model on the task of noise-free image SR in Section~\ref{exp:recon:nonoise}.
For noisy image restoration, we tested the models on realistic image SR, gray-scale image denoising, and color image denoising, in Section~\ref{exp:recon:real}, \ref{exp:recon:gray}, and \ref{exp:recon:color}, respectively, to show the superior performance of the RONet and verify the assumption.

\subsection{Data and implementation details}\label{sec4.1}
We used DIV2K to generate training datasets for all tasks.
DIV2K is a public high-resolution (2k) dataset \cite{ntire/2018}, consisting of 800 training images, 100 validation images, and 100 test images, and ensuring the high quality and diversity of images.
The quality of restored images is evaluated using the standard Peak Signal To Noise Ratio (PSNR) and the Structural Similarity (SSIM) index.


For the task of noise-free image SR, with 4 times at each dimension (x4), we downsampled an HR image from DIV2K using bicubic interpolation to generate a pair of LR and HR images for training. The PSNR and SSIM were measured only on Y channel by ignoring 4 pixels from the boarder for images converted from RGB to YCbCr.
For the task of realistic image SR (x4), we followed the rules of track 2, realistic image SR challenge of NTIRE 2018 \cite{ntire/2018} and used the provided realistic image dataset for training and test. In this track, the LR images were generated via realistic mild adverse conditions, namely, the degradation operators involve the motion blur and the Poisson noise that depend on images, and we have no knowledge about the blur and noise strength. The maximal PSNR and SSIM were measured by cropping a $ 60\times 60 $ patch from the center of an RGB image and shifting it up to 40 pixels in four directions.
For the image denoising task, we added the spatially invariant additive white Gaussian noise (AWGN) to the clean images to generate the training and test data. The PSNR and SSIM were measured for all pixels of an image.

\emph{RODec} is comprised of the cascaded ROPs, and it could be trained by supervised or unsupervised learning. Each branch of ROPs, i.e., \textit{Cblock} or \textit{Rblock}, consists of the three basic blocks whose structures are ``Conv + Relu + Conv'', as shown in Fig \ref{fig:02}. To extract the diverse features, we set the numbers of kernels of these two convolutional layers to be $ 256 $ and $ 64 $, respectively. Besides, we set the number of kernels of the last convolutional layer in ROP to be 3 for color images, and to be 1 for gray-scale images.
For the \emph{unsupervised RODec}, we cropped the patches of size $ 64\times 64 $ from DIV2K, and randomly selected 16 patches to train it in each update.
For the \emph{supervised RODec}, we used the same strategy as the unsupervised case to generate patches and batches, and we obtained the ground truth of a patch by SVD to train the model.

For \emph{RORec}, we adopted the structure of residual networks with the pixel-shuffle strategy, and it was trained supervisedly. The \textit{Upsampling} includes two convolutional layers, whose numbers of kernels are respectively $ 256 $ and $ 3 $. Except for the last layer in \textit{Upsampling}, whose kernel size is $ 9\times 9 $, all the other layers are set to have a kernel size of $ 3\times 3 $.
Each residual block of \textit{RecROs}, \textit{RecRes} and \textit{RecFus} contains two convolutional layers, and their numbers of kernels depend on one particular tasks.
Concretely, for the single image SR,  either noise-free or noisy case, we set the numbers of kernels to be $ 192 $ and $ 48 $ for \textit{RecROs}, and to be $ 256 $ and $ 64 $ for \textit{RecRes} as well as \textit{RecFus}, respectively. Moreover, we set the patch size up to $ 64\times 64 $ to satisfy the requirement of perceptual loss and set the batch size to be 4 to save GPU memory.
For the image denoising, either gray-scale or color mode, we set the numbers of kernels to be $ 96 $ and $ 48 $ for \textit{RecROs}, and to be $ 128 $ and $ 64 $ for \textit{RecRes} as well as \textit{RecFus}, respectively. Besides, we cropped the patches of size $ 64\times 64 $ from DIV2K and used them as ground truth.
We then added AWGN of random noise level ($ \sigma \in \left[ 0, 75\right]  $) to the clean patches, and randomly selected 4 noisy patches to train models in each update.

\begin{table}[!t]
\caption{Evaluation of RO decomposition by the unsupervised RO decomposition (UROD), supervised RO decomposition (SROD) and SVD. The testing results are the average score of 10 images of size $ 1024\times 1024 $, which are cropped from the validation dataset of DIV2K. GPU run time for training is in hours (h) and for test is in seconds (s). \textit{Note that the second and third RO components of SROD are particularly sparse, and the scores therefore are particularly high.}}
\label{tab:RODec}
\resizebox{1\linewidth}{!}{
	\begin{tabular}{cc|ccccc|ccccc}
		\hline
		\multirow{2}{*}{Method}& \multirow{2}{*}{Criterion}&  \multicolumn{4}{c}{RODec for grayscale images}& Runtime & \multicolumn{4}{c}{RODec for RGB images} & Runtime\\
		&  &  $ X_1 $&  $ X_2 $&  $ X_3 $  &  $ E_3 $& Training/Test & $ X_1 $& $ X_2 $&  $ X_3 $  &  $ E_3 $& Training/Test\\
		\hline
		\multirow{2}{*}{UROD}&  PSNR&  38.03&  35.95&  55.79&  18.59& \multirow{2}{*}{6.9h/1.9$ \pm $0.5s}& 42.04&  39.05&  51.66&  18.59& \multirow{2}{*}{7.0h/2.0$ \pm $0.5s}\\
		&  SSIM& 0.99&  0.81&  0.98&  0.28&  &  0.99& 0.91& 0.98&  0.28&\\
		\hline
		\multirow{2}{*}{SROD}&  PSNR&  34.46&  103.3&  127.6&  18.49& \multirow{2}{*}{11.7h/2.1$ \pm $0.8s}&  41.26&  89.57&  99.92&  18.58&  \multirow{2}{*}{18.7h/2.1$ \pm $0.9s}\\
		&  SSIM&  0.99&  1.00&  1.00&  0.27&  & 0.99&  1.00&  1.00&  0.29&\\
		\hline
		\multirow{2}{*}{SVD}&  PSNR&   45.82&  45.57&  45.30&  18.61&  \multirow{2}{*}{--/4.3$ \pm $0.2s}&  45.58&  45.38&  44.91&  18.62&  \multirow{2}{*}{--/12.1$ \pm $0.3s}\\
		&  SSIM& 0.98&  0.98&  0.97&  0.28&  &0.98&  0.97&  0.97&  0.28& \\
		\hline
\end{tabular}}
\end{table}

\begin{figure}[!t]
\centering
\includegraphics[width=1\linewidth]{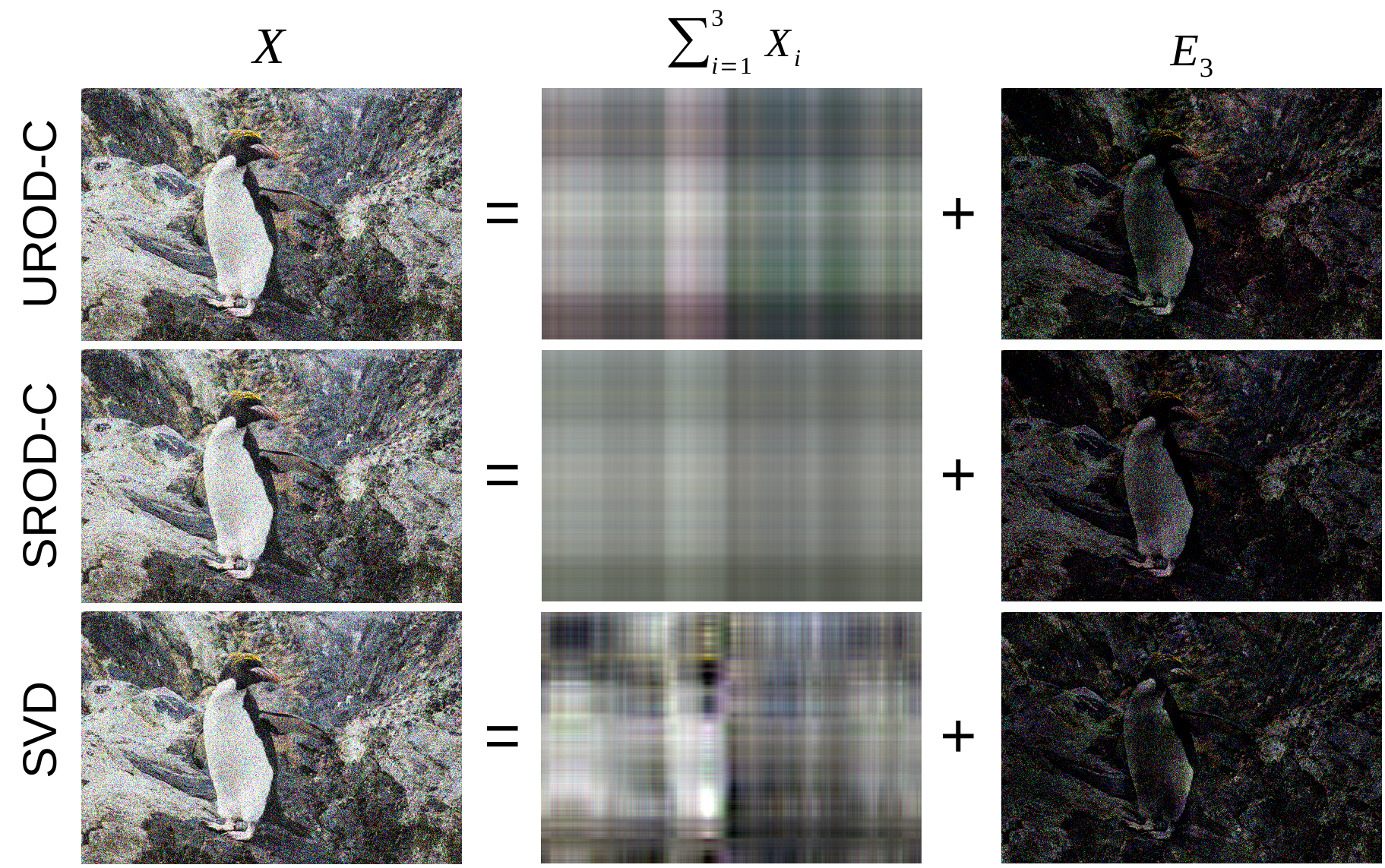}
\caption{Visualization of RO decomposition by UROD-C, SROD-C, and SVD. Here, the noise level of $ X $ is 50.}
\label{fig:rodec}
\end{figure}

The RONet, with RODec and RORec, was trained using the ADAM \cite{Kingma/2015} optimizer  with $ \beta_1 = 0.9 $, $ \beta_2=0.999 $ and $ \epsilon = 1\times 10^{-8} $. 
As RODec is lighter than RORec, we used $ 5\times 10^{5} $ updates to train the former and $ 1\times 10^{6} $ for the latter. The learning rate is initialized to be $ 1\times 10^{-4} $, which decreases to $ 1\times 10^{-5} $ after $ 4\times 10^5 $ updates for RODec, and decays by a factor of 0.5 every $ 2\times 10^5 $ updates for RORec.
Finally, RONet was implemented with TensorFlow \footnote{https://zmiclab.github.io/projects.html}, and all models were trained and tested on a GTX 1080 Ti GPU with 11 GB memory.

\subsection{Performance of RO decomposition}\label{sec:RODec}
In this section, we trained four models which would be used in the following sections. Concretely, we used two anisotropic kernels for ROP units to extract features, i.e., $ 1\times 3 $ for \textit{Cblock} and $ 3\times 1 $ for \textit{Rblock}. Moreover, we set the number of ROP units to 6, i.e., $ L=6 $. For supervised RODec, we adopted the settings shown in Section~\ref{sec4.1} and used the supervised strategy presented in Section~\ref{sec3.4} to train two models for gray-scale and color images, respectively. For convenience, the resulting models are referred to as SROD-G and SROD-C. For unsupervised RODec, we adopted the same settings but utilized the unsupervised approach to train two models for gray-scale and color images, which are referred to as UROD-G and UROD-C, respectively. Note that the number of ROP units could range between $ 1 $ and $ 6 $ in applications for the pre-trained models.

\begin{table*}[!ht]
\centering
\caption{Results of ablation studies: RONet for realistic image SR with different settings, including whether train it in an end-to-end fashion, different RO decompositions, the number of RO projections (\#ROP), the number of residual blocks (\#ResBlocks), and whether use batch normalization (BN). The GPU run time (in hours for training and in seconds for test) and the size of the resulting model (\#Parameters), which denotes the number of parameters from both RO decomposition and reconstruction networks, are reported. The bold font indicates the optimal settings in each of the sub-studies, while the underline font denotes the best settings across the sub-studies. Please refer to the text for details.
}
\label{tab03} \resizebox{1\textwidth}{!}{
	\begin{tabular}{|c||c||c|c||c|c|c|c|c||c|c|c||c|c|}
		\hline
		\multirow{2}{*}{Model}&  \multirow{2}{*}{End-to-end}&  \multicolumn{2}{c||}{RO decomposition}&  \multicolumn{5}{c||}{RO reconstruction}&    \multicolumn{3}{c||}{Evaluation}& \multicolumn{2}{c|}{GPU run time}\\
		\cline{3-14}
		&  &  Method&  \#ROP&  \#ResBlocks&  Loss&  Initializer&  BN&  Scale&  \#Parameters& PSNR& SSIM& Training&  Test\\
		\hline
		\#1& \multirow{3}{*}{N}&  \multirow{3}{*}{UROD-C}&  1& \multirow{3}{*}{$ (3,6,3) $}& \multirow{3}{*}{$ L_2 $}& \multirow{3}{*}{Xavier-U}& \multirow{3}{*}{Y}& \multirow{3}{*}{1.0}&  4.04M&  24.31&  0.5705&  10.8h& 9.4s\\
		\#2& &  &\textbf{3}&  & & & & & 5.04M &  24.33&  0.5727&  17.4h&  10.8s\\
		\#3& &  &  6&  & & & & & 6.53M &  24.21&  0.5607&  26.8h& 12.9s\\
		\hline\hline
		\#4&  \multirow{3}{*}{N}&  \multirow{3}{*}{UROD-C}&  \multirow{3}{*}{3}&  $ (3, 6, 3) $ & \multirow{3}{*}{\underline{$L_1$}}& \multirow{3}{*}{\underline{Xavier-U}}&  \multirow{3}{*}{Y}&  \multirow{3}{*}{1.0}& 5.04M&  24.50&  0.5844&  17.0h&10.7s\\
		\#5&  &  &  &  $ (12, 24, 3) $&  &  &  &  &  11.84M&  24.53&  0.5837&   22.8h& 15.0s\\
		\#6&  &  &  &  $ \mathbf{(8, 16, 8)} $&  &  &  &   &  10.29M&  24.53&  0.5851&  21.2h&  13.9s\\
		\hline\hline
		\#7&  \multirow{3}{*}{N}&  \multirow{3}{*}{UROD-C}&  \multirow{3}{*}{3}&  \multirow{3}{*}{$ (3, 6, 3) $}& \multirow{3}{*}{$ L_1 $}&  \multirow{3}{*}{MSRA-N}&  \textbf{Y}&  \textbf{1.0}&  5.04M&  24.46&  0.5814&  17.4h& 10.7s\\
		\#8&  &  &  &  &  &  & N&  1.0&  5.04M&  24.46&  0.5802&  17.0h& 10.3s\\
		\#9&  &  &  &  &  &  & Y&  0.1&  5.04M&  24.42&  0.5788&   17.6h& 10.6s\\
		\hline\hline
		\#10& N&  SROD-C& 3&  $ (3, 6, 3) $& $ L_1 $&  Xavier-U&  Y&  1.0& 5.04M& 24.50& 0.5835& 17.4h& 10.6s\\
		\hline
		\#11& N&  SVD& --&  $ (3, 6, 3) $& $ L_1 $&  Xavier-U&  Y&  1.0&  3.7M&  24.46& 0.5803& 99.4h& 12.9s\\
		\hline
		\#12& Y&  UROD-C& 3&  $ (3, 6, 3) $& $ L_1 $&  Xavier-U&  Y&  1.0& 5.04M& 24.47& 0.5844& 39.5h& 10.7s\\
		\hline
\end{tabular}}
\end{table*}

To study the performance of RODec in extracting RO components, we tested the four models on ten $ 1024\times 1024 $ images cropped from DIV2K. Concretely, we first added AWGN with the noise level of 30 to the ten clean images. Then, we used UROD to extract the RO components and the residual error of a noisy image, i.e., $ X_1, X_2, X_3 $ and $ E_3 $, and compared them with the corresponding components of the ground truth, which were also extracted via UROD. Similarly, we repeated that for SROD and SVD, and reported the performance of them in Table \ref{tab:RODec}. The PSNR and SSIM of residual error are particularly low for each method, which means that most of the added noise is included in the residual component, and the RO components therefore are almost noise-free. The average test runtime of UROD and SROD are about one-sixth of that of SVD, which demonstrates that the proposed RO decomposition is much faster than SVD in extracting the first three RO components. Fig. \ref{fig:rodec} visualizes the performance of UROD, SROD, and SVD in decomposing a color image, which shows that each of them could extract a low-rank component (the sum of the first three RO components) that is almost noise-free. However, the low-rank component extracted by SVD seems more informative, which demonstrates that it is necessary to explore more robust DNNs-based methods for RO decomposition.

\subsection{Ablation studies}\label{sec43}

In this study, we trained twelve RONet models and tested them for realistic image SR using the noisy and blurry images from DIV2K. The parameters include the settings for RODec, RORec, and the training schemes of initializations, batch normalization, residual scaling, and the methods of RO decomposition.
The parameter settings and results are summarized in Table~\ref{tab03}. \textit{Note that we used total $ 2\times 10^5 $ updates to train each model. The learning rate started from $ 1\times 10^{-4} $ and decreased to $ 1\times 10^{-5} $ after $ 1\times 10^5 $ updates.}

To study the effect of ROP units, we trained and compared three models.
We first adopted the pre-trained UROD-C as the RO decomposition method and set the number of ROPs to be different integers, i.e., 1, 3, and 6. We further fixed the parameters of UROD-C and trained the RORec by optimizing the loss function in (\ref{eq15}), where, $ \alpha=2 $, $ \lambda=0.5 $, and $ \eta=1\times 10^{-3} $.
Table~\ref{tab03} presents the detailed settings and results:
model \#2, with 3 ROPs, delivered the best performance.
Therefore, we adopted these settings of RODec for the models in the following studies.

To study RORec, we further trained and compared three models.
We first studied the effect of different loss functions, namely we changed $ \alpha $ to be 1 for model \#4.
The result shows that $L_1$ performed better than $L_2$ used by model \#2.
We then studied the effect of RORec with different numbers of residual blocks in \textit{RecROs}, \textit{RecRes} and \textit{RecFus}.
Table \ref{tab03} shows that the three models achieved comparable mean PSNR and SSIM values in restoration,
but model \#4 is lighter and computationally more efficient.

To study the training schemes, we trained another three models.
We compared the two initialization methods, i.e., Xavier uniform approach \cite{He_delving/2015} and MASRA normal method \cite{Glorot/2010}, and found the former delivered better accuracies.
We studied the batch normalization (BN) strategy, by comparing model \#7 and model \#8, and found BN could improve the performance.
Finally, we multiplied 0.1 to the residuals before adding them to the identity maps in the residual blocks for model \#9, to study the effect of residual scaling.
The results in Table~\ref{tab03} show that model \#7 without the scaling is better than model \#9 with this scaling in terms of accuracy.

To study the methods of RO decomposition, we trained two models using different RO decompositions. We first replaced the UROD-C with the pre-trained SROD-C and maintained other settings as model \#4 used to obtain model \#10. We further used the conventional SVD to achieve RO decomposition and studied its efficiency. 
Note that the computation time of SVD-based RO decomposition is about six times of that of SROD-C, we therefore limited the training procedure to be within an acceptable time. To study the effect of training RONet in an end-to-end fashion, we first adopted the settings as model \#4 used and randomly initialized the parameters of RODec and RORec. After that, we jointly trained them by minimizing $ L^{unsup}_{Dec} + L_{Rec} $ and obtained an end-to-end model, i.e., model \#12. The comparisons between model \#4 and model \#12 showed that training RONet in an end-to-end fashion increased its training time, but did not improve its performance. Moreover, the results showed model \#4 could deliver better performance than model \#10, we therefore adopted the unsupervised RODec in the following tasks.

\subsection{Performance of Rank-One Network}\label{exp:recon}
This section studies the applications of the proposed RONet framework for four image restoration tasks, i.e., noise-free image SR, realistic image SR, gray-scale image denoising and color image denoising.
We set the number of ROP to be 3 for image SR, according to the ablation study results in Table \ref{tab03}, and to be 1 for image denoising to decrease computational complexity.
For each task, we trained a task-specific RORec model, which has the same architecture and setting as the RORec in model \#4. 
We adopted the pre-trained unsupervised RODec models, as shown in Section 4.2, and fixed their parameters when we trained the RORec models. Moreover, we reported the total number of parameters of RONet, \textit{which is comprised of the parameters from RODec and RORec,} in the following sections. Note that we obtained the restored images, based on which we computed their PSNR and SSIM, using the released codes and pre-trained models for the compared methods.

\begin{table*}[!t]
\centering
\caption{Results of noise-free image SR: including SRCNN \cite{Dong/2016}, EDSR \cite{Lim/2017}, EnhanceNet \cite{Sajjadi/2017}, SRGAN \cite{Ledig/2017}, ESRGAN \cite{Wang/2018}, OISR \cite{He/OISR/2019}, and RONet. The sizes of resulting models are given in the row of \#Parameters.
	The bold and italic fonts respectively denote the best and second-best results in each row.
	\emph{Here, RONet-NF does not demonstrate particularly better performance,
		because its function of separating noise from RO components of the image may not be useful when super-resolving noise-free images.} Note that EnhanceNet, SRGAN, and ESRGAN are the perceptual quality-oriented methods, so they do not achieve high scores, but they could reconstruct visually more realistic images, as shown in Fig \ref{fig:05}.
}
\label{tab02}
\resizebox{1\textwidth}{!}{
	\begin{tabular}{|c|c|c|c|c|c|c|c|c|c|}
		\hline
		Datasets&  Criterion&  Bicubic&  SRCNN&  EDSR&  EnhanceNet&  SRGAN&  ESRGAN& OISR-RK2& RONet-NF\\
		\hline
		\multirow{2}{*}{Set5}&  PSNR&  28.42&  30.49&  \textbf{32.46}&  28.56&  28.19&  30.44&  \textit{32.33}&  31.86\\
		&  SSIM&  0.8105&  0.8629&  \textbf{0.8976}&  0.8093&  0.8163&  0.8505&  \textit{0.8958}&  0.8894\\
		\hline
		\multirow{2}{*}{Set14}&  PSNR&  26.10&  27.61&  \textbf{28.80}&  25.04&  25.97&  26.28&  \textit{28.71}&  28.42\\
		&  SSIM&  0.7048&  0.7535&  \textbf{0.7872}&  0.6528&  0.7001&  0.6974&  \textit{0.7844}&  0.7767\\
		\hline
		\multirow{2}{*}{BSD100}&  PSNR&  25.96&  26.91&  \textbf{27.72}&  24.09&  24.63&  25.30&  \textit{27.65}&  27.44\\
		&  SSIM&  0.6676&  0.7104&  \textbf{0.7414}&  0.6006&  0.6416&  0.6494&  \textit{0.7386}&  0.7313\\
		\hline
		\multirow{2}{*}{Urban100}&  PSNR&  23.15&  24.53&  \textbf{26.64}&  22.30&  23.67&  24.35& \textit{26.37}&  25.73\\
		&  SSIM&  0.6579&  0.7230& \textbf{0.8029}&  0.6504&  0.6984&  0.7322&  \textit{0.7952}&  0.7739\\
		\hline\hline
		\multicolumn{2}{|c|}{\# Parameters of models} &  --&  0.07M&  43.1M&  0.85M&  0.78M&  16.6M&  5.5M&  5.0M\\
		\hline
\end{tabular}}

\end{table*}
\begin{figure*}[h]
\centering
\includegraphics[width=1\linewidth]{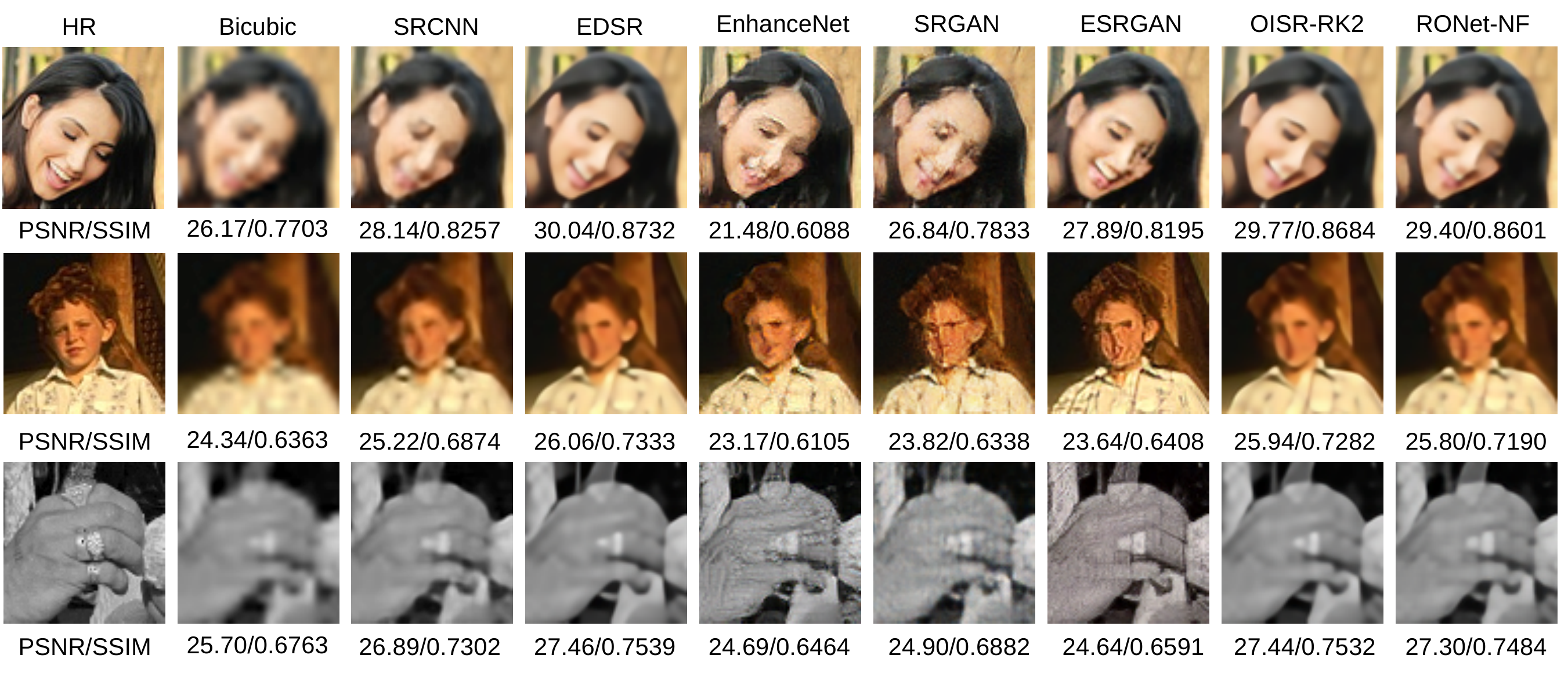}\\[-2ex]
\caption{Visualization of noise-free image SR: three typical examples (cropped patches of size $ 80\times 80 $) from the four datasets. Please refer to Supplementary Material for high-resolution images.
}
\label{fig:05}
\end{figure*}

\subsubsection{Noise-free image super-resolution}\label{exp:recon:nonoise}

To simulate the noise-free case, we generated the LR images by downsampling the HR images in DIV2K using bicubic interpolation.
The resulting LR and HR image pairs were then used for the training of the SR models.
For convenience, the proposed RONet model for this noise-free image SR is referred to as RONet-NF.
We evaluated the performance of RONet-NF on four public datasets, i.e., Set5 \cite{Bevilacqua/2012}, Set14 \cite{Zeyde/2012}, BSD100 \cite{Martin/2001}, and Urban100 \cite{Huang/2015}, and compared with SRCNN \cite{Dong/2016}, EnhanceNet \cite{Sajjadi/2017}, SRGAN \cite{Ledig/2017}, ESRGAN \cite{Wang/2018}, EDSR \cite{Lim/2017}, and OISR \cite{He/OISR/2019}, which are the state-of-the-art methods for noise-free image SR.

We only trained the reconstruction module (RORec) of RONet-NF, namely, we adopted the pre-trained UROD-C to achieve RO decomposition, and fixed its weights to update the parameters of RORec.  For the loss function, we set $ \alpha $ in (\ref{eq15}) to be 1, since the previous works show that $ L_1 $ regularization could achieve better performance than $ L_2 $; $ \eta $ in (\ref{eq15}) was set to be $ 1\times 10^{-3} $, and $ \lambda $ in (\ref{eq15}) was set to be $ 0.5 $.

To be consistent with the previous methods which were evaluated on YCbCr images, we converted the mode of restored images to be YCbCr and computed the PSRN and SSIM only on the Y channel, as shown in Section \ref{sec4.1}. We reported the quantitative evaluations of RONet-NF and six state-of-the-art methods in Table \ref{tab02}, which shows that EDSR achieves the highest PSNR and SSIM, OISR-RK2 ranks the second-best, and RONet-NF delivers the third-best performance.
We visualized the restored results of the typical cases in Fig \ref{fig:05}. One can see that the images super-resolved by EnhanceNet, SRGAN, and ESRGAN tend to have artifacts, but are visually more realistic.
It should be noted that the noise-free image SR task does not require the function to separate the RO components from the noise of images, which is the major advantage of the proposed RONet framework.
Hence, RONet-NF did not demonstrate particularly better performance, compared to the other state-of-the-art methods.

\begin{table}[!t]
\centering
\caption{Results of realistic image SR: including EDSR \cite{Lim/2017}, WDSR \cite{Yu/2018} and RONet-R.
	The bold font denotes the best PSNR or SSIM.
	\emph{Here, RONet-R achieves particularly better PSNR and SSIM values than the others. This is because its function of separating noise from RO components of the images could play an important role when super-resolving real images.}
} \label{tab04}
\resizebox{1\linewidth}{!}{
	\begin{tabular}{ccccccccc}
		\hline
		\multirow{2}{*}{Criterion}&  \multirow{2}{*}{Bicubic}&  \multirow{2}{*}{EDSR}&  \multirow{2}{*}{WDSR}& \multicolumn{5}{c}{RONet-R by different settings of $ \lambda $}\\
		\cmidrule(lr){5-9}
		&  &  &  &  0&  0.001&  0.01&  0.1&  0.5\\
		\hline
		PSNR&  23.16&  24.40&  24.42&   24.50&  \textbf{24.53}&  \textit{24.52}&  24.51& 24.51 \\
		SSIM&  0.5178&  0.5796&  0.5833&   0.5854& \textbf{0.5868}& \textit{0.5863}& 0.5855&  0.5848\\
		\# Paras&  --&  43.1M&  9.5M&  5.0M& 5.0M&  5.0M&  5.0M&  5.0M\\
		\hline
\end{tabular}}
\end{table}

\begin{figure}[!ht]
\centering
\includegraphics[width=0.8\linewidth]{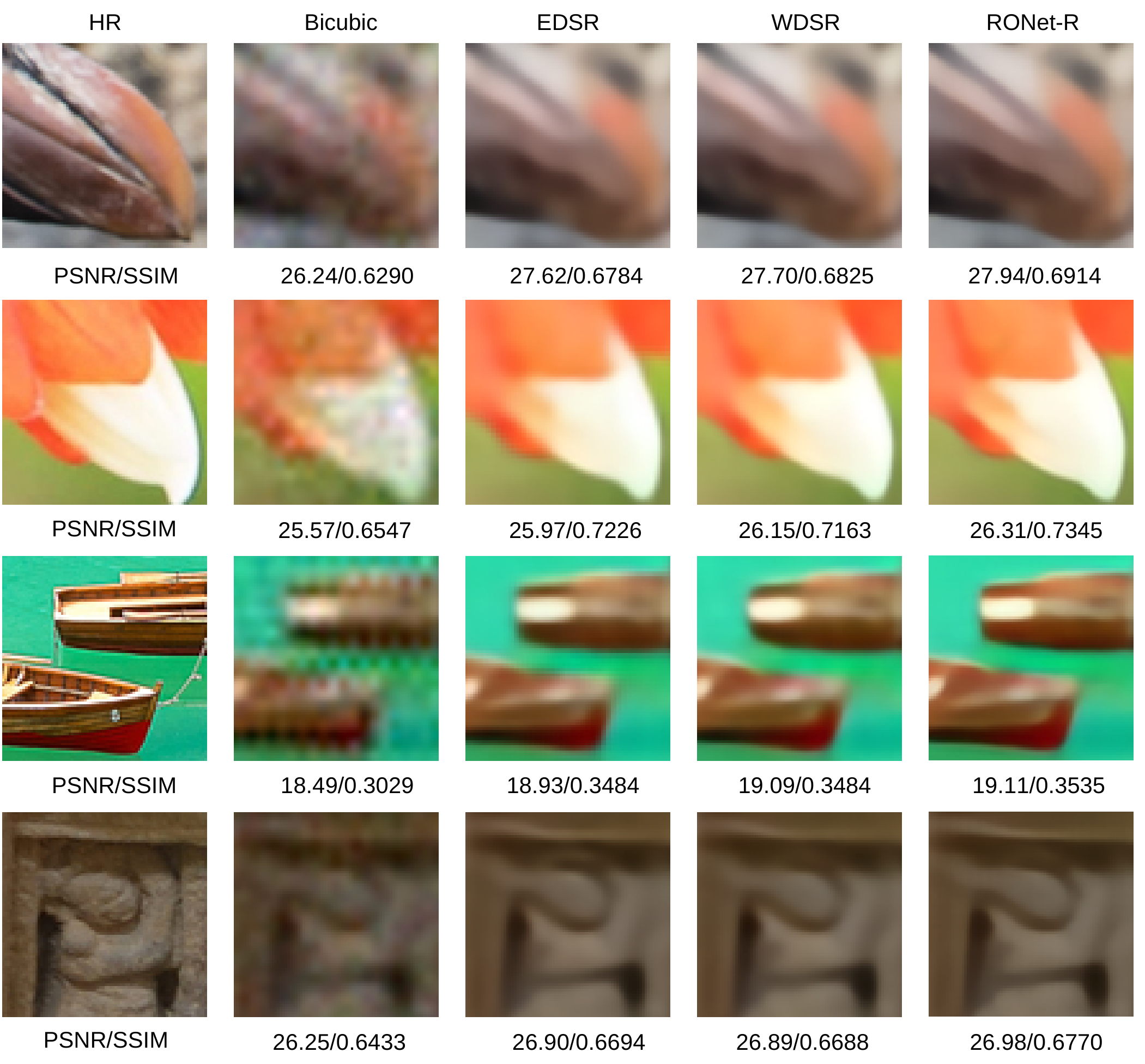}\\[-2ex]
\caption{Visualization of realistic image SR: four typical examples (cropped patches of size $ 80\times 80 $). Please refer to Supplementary Material for high-resolution images.
}
\label{fig:06}
\end{figure}

\subsubsection{Realistic image super-resolution}\label{exp:recon:real}

\begin{figure*}[!ht]
\centering
\includegraphics[width=1\linewidth]{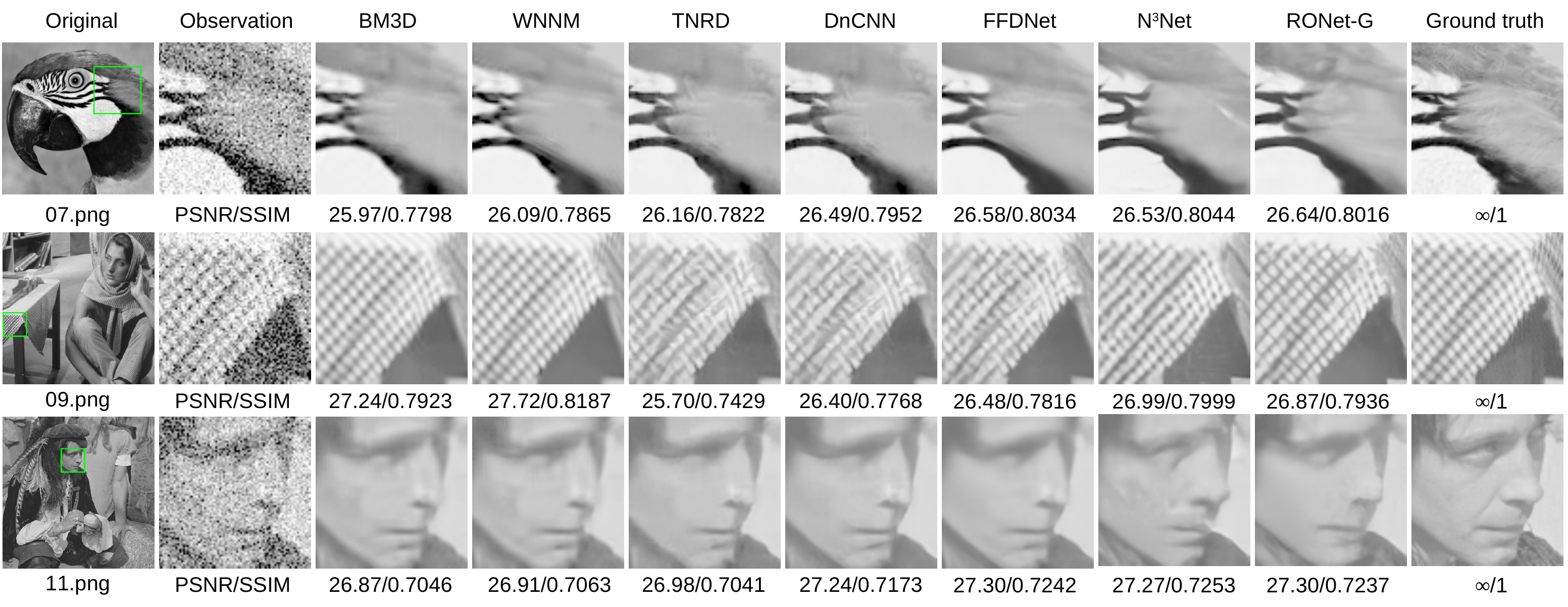}\\[-2ex]
\caption{Visualization of gray-scale image denoising: three typical examples (cropped patches of size $ 80\times 80 $) from Set12. Please refer to Supplementary Material for high-resolution images.}
\label{fig:07}
\end{figure*}
\begin{table}[h]
\centering
\caption{The quantitative evaluation of BM3D \cite{Dabov/2007}, WNNW \cite{Gu/2014}, TNRD \cite{Chen/2016}, DnCNN \cite{Zhang_beyond/2017}, FFDNet \cite{Zhang_ffdnet/2018}, N$ ^3 $Net \cite{Ploetz/2018}, and RONet-G on the task of gray-scale image denoising. We tested these methods on RNI6 \cite{Lebrun/2015}, Set12, and BSD68 \cite{Roth/2005} with different noise levels, i.e., $ \sigma=15, 25, 35, 50 $, and reported their performance. The bold value denotes the best performance, while the italic value represents the second best performance. }
\label{tab06}
\resizebox{1\linewidth}{!}{
	\begin{tabular}{ccccccccccc}
		\hline
		\multirow{2}{*}{Dataset}&  \multirow{2}{*}{Method}&  \multicolumn{2}{c}{$ \sigma=15 $}&  \multicolumn{2}{c}{$ \sigma=25 $}&  \multicolumn{2}{c}{$ \sigma=35 $}&  \multicolumn{2}{c}{$ \sigma=50 $}&\multirow{2}{*}{\#Paras}\\
		\cmidrule(lr){3-4}\cmidrule(lr){5-6}\cmidrule(lr){7-8}\cmidrule(lr){9-10}
		&  &  PSNR&  SSIM&  PSNR&  SSIM&  PSNR&  SSIM&  PSNR&  SSIM& \\
		\hline
		\multirow{8}{*}{RNI6}&  BM3D& 30.23& 0.8383& 27.98& 0.7626& 26.61& 0.7156& 25.32& 0.6691& --\\
		&  WNNW& 30.40& 0.8424& 28.21& 0.7707& 26.87& 0.7209& 25.49& 0.6760& --\\
		&  TNRD& 30.27& 0.8472& 28.02& 0.7663& --& --& 25.33& 0.6688& 0.58M\\
		&  DnCNN& 30.02& 0.8266& 28.07& 0.7568& 26.90& 0.7172& 25.63& 0.6757& 0.66M\\
		&  FFDNet& 30.43& 0.8561& 28.27& 0.7801& 27.01& 0.7298& \textit{25.74}& \textit{0.6843}& 0.48M\\
		&  N$ ^3 $Net& \textbf{30.55}& \textbf{0.8604}& \textbf{28.33}& \textbf{0.7835}& \textbf{27.09}& \textbf{0.7352}& \textbf{25.79}& \textbf{0.6865}& 0.71M\\
		&  RONet-G& \textit{30.51}& \textit{0.8577}& \textit{28.31}& \textit{0.7811}& \textit{27.02}& \textit{0.7302}& 25.73& 0.6839& 2.01M\\
		\hline
		\multirow{8}{*}{Set12}&  BM3D& 32.40& 0.8985& 29.99& 0.8520& 28.41& 0.8120& 26.79& 0.7644& --\\
		&  WNNW& 32.70& 0.9002& 30.23& 0.8571& 28.66& 0.8196& 26.97& 0.7745& --\\
		&  TNRD& 32.50& 0.8987& 30.05& 0.8528& --& --& 26.81& 0.7634& 0.58M\\
		&  DnCNN& 32.68& 0.9027& 30.36& 0.8618& 28.83& 0.8263& 27.21& 0.7806& 0.66M\\
		&  FFDNet& 32.75& 0.9055& 30.43& 0.8653& 28.92& 0.8316& 27.32& \textit{0.7888}& 0.48M\\
		&  N$ ^3 $Net& \textit{32.91}& \textbf{0.9073}& \textit{30.50}& \textit{0.8660}& \textbf{29.03}& \textbf{0.8339}& \textbf{27.44}& \textbf{0.7926}& 0.71M\\
		&  RONet-G& \textbf{32.92}& \textit{0.9066}& \textbf{30.54}& \textbf{0.8661}& \textit{28.98}& \textit{0.8318}& \textit{27.37}& 0.7881& 2.01M\\
		\hline
		\multirow{7}{*}{BSD68}& BM3D& 31.12& 0.8789& 28.62& 0.8093& 27.13& 0.7548& 25.73& 0.6945& --\\
		&  WNNW& 31.32& 0.8813& 28.84& 0.8183& 27.32& 0.7634& 25.85& 0.7005& --\\
		&  TNRD& 31.42& 0.8870& 28.92& 0.8201& --& --& 25.97& 0.7032& 0.58M\\
		&  DnCNN& 31.61& 0.8915& 29.16& 0.8293& 27.68& 0.7784& 26.23& 0.7180& 0.66M\\
		&  FFDNet& 31.63& 0.8951& 29.19& 0.8342& 27.73& 0.7848& 26.29& \textit{0.7270}& 0.48M\\
		&  N$ ^3 $Net& \textbf{31.79}& \textbf{0.8979}& \textbf{29.29}& \textbf{0.8373}& \textbf{27.83}& \textbf{0.7894}& \textbf{26.38}& \textbf{0.7317}& 0.71M\\
		&  RONet-G& \textit{31.77}& \textit{0.8969}& \textit{29.28}& \textit{0.8356}& \textit{27.79}& \textit{0.7853}& \textit{26.34}& 0.7269& 2.01M\\
		\hline
\end{tabular}}
\end{table}

To study the performance of RONet in super-resolving realistic images, we trained the model using the blurry and noisy images of DIV2K from the NTIRE2018 image SR challenge \cite{ntire/2018}. We adopted the pre-trained UROD-C to achieve the RO decomposition and obeyed the settings as model \#4 used in Table~\ref{tab03} for RORec. We set $ \alpha=1 $, $ \lambda=0.5 $, as well as $ \eta=1\times 10^{-3} $ and used the training strategy in Section \ref{sec3.4} to update the parameters of RORec up to $ 8\times 10^5 $ times. To study the effect of $ \lambda $, we first set $ \lambda=0.001,0.01,0.1,0.5 $ as well as $\lambda=0$. We further fine-tune the pre-trained RORec using extra $ 2\times 10^5 $ updates for each setting.
For convenience, the resulting models are referred to as RONet-R. We compared RONet-R with EDSR \cite{Lim/2017} and WDSR \cite{Yu/2018}, which were the winners of NTIRE 2017 and 2018 image SR challenges, respectively.
All models were evaluated using the validation dataset (100 images) of the realistic DIV2K,
and we used the same strategy from NTIRE challenges to compute the PSNR and SSIM, as shown in Section \ref{sec4.1}.

We reported the quantitative evaluations of RONet-R and two state-of-the-art methods in
Table \ref{tab04}, where sizes of the resulting network models are also given.
RONet-R is the lightest model, compared to EDSR and WDSR, though it achieved the best PSNR and SSIM.
Moreover, RONet-R with $ \lambda=0.001 $ or $ 0.01 $ achieved better performance than that with $\lambda=0$, and RONet-R with $ \lambda=0.001 $ were the best in this study. This indicates proper deep supervision could improve the performance in restoration, though RONet-R with $ \lambda=0,0.1,0.5 $ achieved similar performance.
Fig \ref{fig:06} visualizes the restored images of four typical cases.
One can see that RONet-R tends to generate fine details, whereas EDSR and WDSR are prone to smooth them.
We conclude that in realistic image SR, RONet-R could perform particularly well, thanks to its function of separating noise from RO components of the images when restoring them.

\subsubsection{Gray-scale image denoising}\label{exp:recon:gray}
To study the performance of RONet in gray-scale image denoising, we generated a set of noisy images for training.
We converted each clean image from DIV2K to be gray-scale mode and added an AWGN with known noise strength, i.e., $ \sigma=15,25,35 $ or $ 50 $, to generate a noisy observation. We used the mean squared error (MSE) as the loss function, namely we set $ \alpha=2$, $ \lambda=0 $ and $ \eta =0 $ for the hyper-parameters in (\ref{eq15}). We adopted the pre-trained UROD-G to achieve the RO decomposition, and obeyed other settings as model \#1 used in Table \ref{tab03} for RORec, and adopted the training strategy in Section \ref{sec4.1}.
We adopted the settings of kernels for image denoising, as shown in Section \ref{sec4.1}, and trained a model referred to as RONet-G, which was compared with six state-of-the-art methods for gray image denoising,
i.e., BM3D \cite{Dabov/2007}, WNNW \cite{Gu/2014}, TNRD \cite{Chen/2016}, DnCNN \cite{Zhang_beyond/2017}, FFDNet \cite{Zhang_ffdnet/2018}, and N$ ^3 $Net \cite{Ploetz/2018}.
We evaluated their performance in gray-scale image denoising by three widely used datasets, i.e., RNI6 \cite{Lebrun/2015}, Set12, and BSD68 \cite{Roth/2005}.

Table \ref{tab06} summarizes the results of gray-scale image denoising.
Although N$ ^3 $Net achieved overall higher scores than RONet-G, their performance was comparable in terms of PSNR and SSIM. To visually compare these methods, we showed three typical examples in Fig \ref{fig:07}. One can see that WNNM and BM3D could generate more realistic textures for \texttt{09.png}, and RONet-G is prone to reconstruct more details corrupted by noise for \texttt{07.png} and \texttt{11.png}.

\subsubsection{Color image denoising}\label{exp:recon:color}

To study the performance of RONet in color image denoising, we generated a set of noisy images for training.
We added an AWGN to each clean image from DIV2K, with noise levels ranging between $ \left[0, 75 \right] $, to generate a noisy observation. We first adopted the pre-trained UROD-C to achieve the RO decomposition. We further set up a \textit{baseline} by setting the number of kernels to be 48 for all convolutional layers of the residual blocks in \textit{RecROs}, and to be 64 for that in \textit{RecRes} as well as \textit{RecFus}. Moreover, we adopted the original settings of kernels for image denoising, as shown in Section \ref{sec4.1}, and trained another model referred to as RONet-C.
Both of them were compared with three state-of-the-art methods for color image denoising,
i.e., CBM3D \cite{Dabov/2007}, CDnCNN \cite{Zhang_beyond/2017} and FFDNet \cite{Zhang_ffdnet/2018}.
We evaluated the performance of image restoration using three public datasets, i.e., CBSD68 \cite{Roth/2005}, Kodak24 \cite{Franzen/1999} and McMaster \cite{Zhang_color/2011}. Note that noise levels should be given for CBM3D, therefore we evaluated all methods for specific noise levels, i.e., $\sigma=15, 25, 35, 50$.

\begin{figure}[!t]
\centering
\includegraphics[width=1\linewidth]{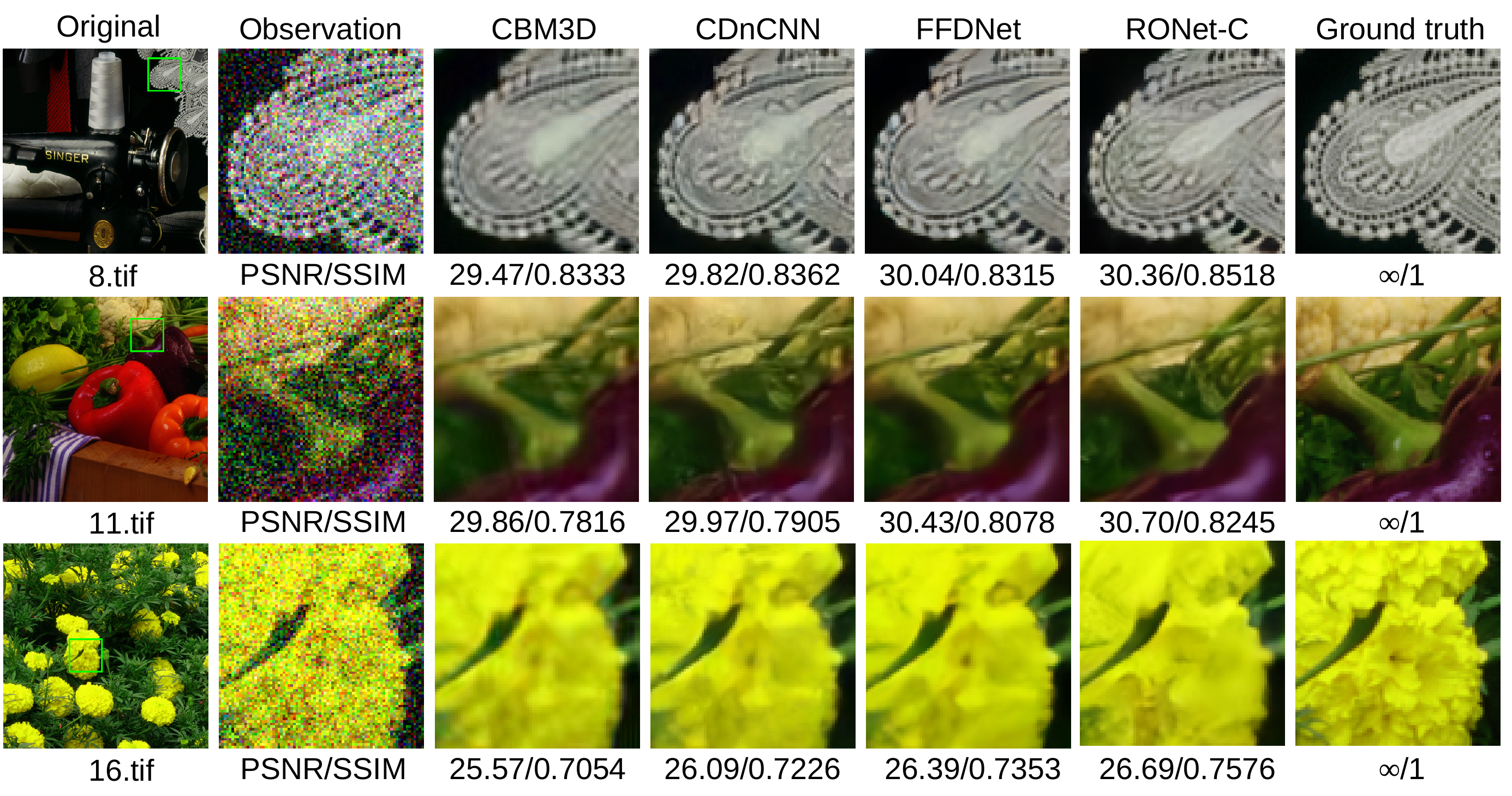}\\[-2ex]
\caption{Visualization of color image denoising: three typical examples (cropped patches of size $ 80\times 80 $) from McMaster. Please refer to Supplementary Material for high-resolution images.}
\label{fig:08}
\end{figure}

\begin{table}[!h]
\centering
\caption{Results of color image denoising: CBM3D \cite{Dabov/2007}, CDnCNN \cite{Zhang_beyond/2017}, FFDNet \cite{Zhang_ffdnet/2018}, and RONet-C. The methods are evaluated on CBSD68 \cite{Roth/2005}, Kodak24 \cite{Franzen/1999}, and McMaster \cite{Zhang_color/2011} with four noise levels.
	The bold and italic fonts respectively denote the best and second-best results in each group.
}
\label{tab05}
\resizebox{1\linewidth}{!}{
	\begin{tabular}{ccccccccccc}
		\hline
		\multirow{2}{*}{Dataset}&  \multirow{2}{*}{Method}&  \multicolumn{2}{c}{$ \sigma=15 $}&  \multicolumn{2}{c}{$ \sigma=25 $}&  \multicolumn{2}{c}{$ \sigma=35 $}&  \multicolumn{2}{c}{$ \sigma=50 $}&\multirow{2}{*}{\#Paras}\\
		\cmidrule(lr){3-4}\cmidrule(lr){5-6}\cmidrule(lr){7-8}\cmidrule(lr){9-10}
		&  &  PSNR&  SSIM&  PSNR&  SSIM&  PSNR&  SSIM&  PSNR&  SSIM& \\
		\hline
		\multirow{4}{*}{CBSD68}&  CBM3D&   33.52&   0.9248&   30.71&   0.8716&   28.89&   0.8207&    27.38&    0.7669 & --\\
		&  CDnCNN&   33.89&   0.9317&   31.23&   0.8863&  29.58&  0.8452&    27.92&    0.7915 & 0.67M\\
		&  FFDNet&   33.87&   0.9318&   31.21&   0.8857&   29.58&   0.8445&    27.96&     0.7916 & 0.83M\\
		&  Baseline&   \textit{33.94}&   \textit{0.9330}&   \textit{31.32}&   \textit{0.8895}&   \textit{29.71}&  \textit{0.8506}&    \textit{28.10}&    \textit{0.8001}& 1.24M\\
		&  RONet-C&   \textbf{33.99}&   \textbf{0.9336}&   \textbf{31.36}&   \textbf{0.8902}&   \textbf{29.74}&  \textbf{0.8514}&    \textbf{28.14}&    \textbf{0.8009}& 2.03M\\
		\hline
		\multirow{4}{*}{Kodak24}&  CBM3D&   34.28&   0.9164&   31.68&   0.8682&   29.90&   0.8212&    28.46&    0.7751& --\\
		&  CDnCNN&   34.48&   0.9212&   32.03&   0.8774&   30.46&   0.8390&    28.85&     0.7895& 0.67M\\
		&  FFDNet&   34.64&   0.9230&  32.13&   0.8790&   30.57&   0.8407&    28.98&    0.7929& 0.83M\\
		&  Baseline&   \textit{34.73}&   \textit{0.9246}&   \textit{32.27}&   \textit{0.8833}&   \textit{30.71}&  \textit{0.8470}&    \textit{29.13}&    \textit{0.8005}& 1.24M\\
		&  RONet-C&   \textbf{34.80}&   \textbf{0.9254}&   \textbf{32.33}&   \textbf{0.8845}&   \textbf{30.77}&   \textbf{0.8484}&    \textbf{29.18}&    \textbf{0.8020}& 2.03M\\
		\hline
		\multirow{4}{*}{McMaster}&  CBM3D&   34.06&   0.9150&   31.66&   0.8739&   29.92&   0.8327&    28.51&    0.7934& --\\
		&  CDnCNN&   33.44&   0.9070&   31.51&   0.8724&   30.14&   0.8412&    28.61&     0.7985& 0.67M\\
		&  FFDNet&   34.66&   \textit{0.9247}&   32.35&   0.8891&   30.81&   0.8573&    29.18&   0.8157& 0.83M\\
		&  Baseline&   \textit{34.68}&   0.9236&   \textit{32.44}&   \textit{0.8898}&   \textit{30.94}&  \textit{0.8598}&    \textit{29.33}&    \textit{0.8203}& 1.24M\\
		&  RONet-C&  \textbf{34.77}&   \textbf{0.9251}&   \textbf{32.51}&   \textbf{0.8920}&   \textbf{31.00}&   \textbf{0.8627}&    \textbf{29.39}&     \textbf{0.8245}& 2.03M\\
		\hline
\end{tabular}}
\end{table}

Table \ref{tab05} summarizes the results of image denoising.
One can see that RONet-C achieved the best scores in all studies, and the baseline got the second-best scores in 23 studies (out of 24).
This was probably due to the advantageous function of separating noise from RO components of the RONet, which plays an important role in restoring images with noise.
Finally, Fig \ref{fig:08} provides three typical examples. Similar to the results of realistic image SR and gray-scale image denoising, RONet-C demonstrated the potential of reconstructing more details corrupted by noise.

\begin{figure}[!ht]
\centering
\includegraphics[width=0.9\linewidth]{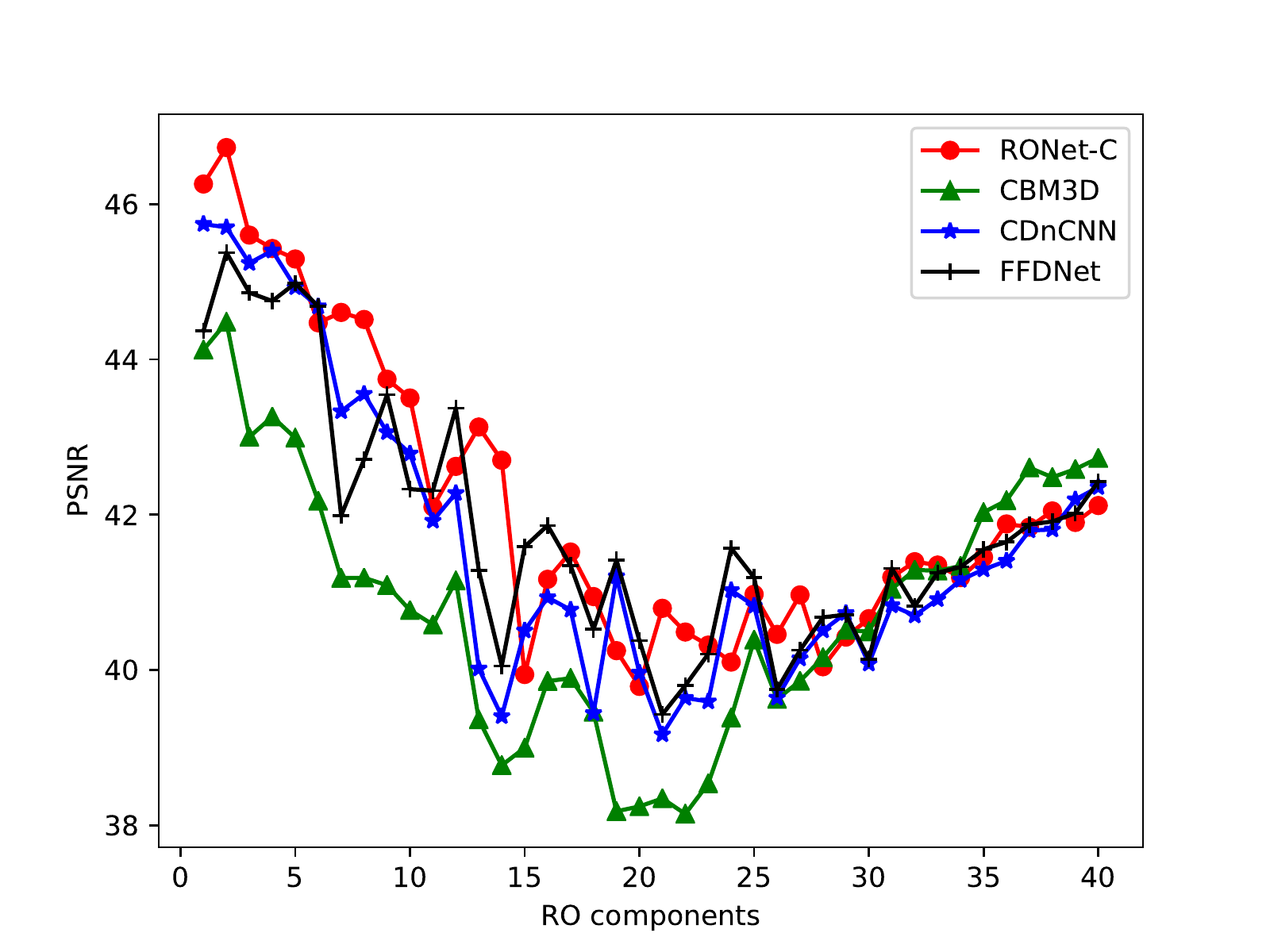}
\caption{PSNR of the RO components for CBM3D, CDnCNN, FFDNet, and RONet-C when the noise level is $ 50 $.}
\label{fig:PSNRRO}
\end{figure}

To demonstrate the advantage of RONet in reconstructing RO components (by RecROs) and residual (by RecRes), we compared the SVD of restored images with the SVD of their ground truth, and plotted the average PSNR of the $ i $-th ($ i=1,\dots, 40 $) RO components, which were obtained from the reconstructed images. Fig~\ref{fig:PSNRRO} shows the results. One can see that RONet achieves the best scores on the first 10 RO components except for the sixth. This demonstrates that RONet decreases the corruption of principal RO components by processing the RO components and residual separately, and thus improves the performance of denoising.

\section{Conclusion}\label{sec06}

In this paper, we formulate a new framework for image restoration based on RO decomposition and reconstruction.
The method separates the RO components of a corrupted image and utilizes the components and residual error to restore the image. To achieve this, we first develop an RO decomposition network to extract the RO components. This network functions by cascading the proposed RO projection. Then, we process the RO components and residual error separately and restore the image from the concatenation of them using the RO reconstruction network.
The implementation of the proposed method based on neural networks is referred to as RONet.

In the experiments, we evaluated RONet using four tasks, i.e., noise-free image SR, realistic image SR, gray-scale image denoising, and color image denoising.
The quantitative and qualitative results showed that RONet was effective and efficient for image restoration.
Particularly, when the images had been degraded with noise and artifacts, RONet demonstrated superior performance.

Regarding to the future work, methodologies for efficient and accurate extraction of RO components could be further explored.
Currently, a fixed number of RO projections, i.e., ranging between 1 and 6, was used for the RODec, according to the ablation study.
However, in reality this number could be adaptive, based on the fact that the ranks could vary greatly across different images.
Moreover, exactly estimating the rank of a realistic image is difficult, and it is still an open question in low-rank matrix completion \cite{Candes_matrix/2010}.
Hence, future work could include the investigation of alternatives for RO decomposition and the exploration of sparsity as well as low-rank property of images.

\section*{Acknowledgment}
This work was funded by the National Natural Science Foundation of China (NSFC) Grant (61971142).
The authors would like to thank Lei Li, Fuping Wu and Xinzhe Luo for useful comments and proof read of the manuscript.

\bibliographystyle{abbrv}
\bibliography{references.bib}
\end{document}